\documentclass[fleqn,usenatbib]{mnras}

\usepackage{newtxtext,newtxmath}

\usepackage[T1]{fontenc}
\usepackage{subcaption}
\usepackage{adjustbox}
\usepackage{threeparttable}
\usepackage{comment}

\DeclareRobustCommand{\VAN}[3]{#2}
\let\VANthebibliography\thebibliography
\def\thebibliography{\DeclareRobustCommand{\VAN}[3]{##3}\VANthebibliography}

\usepackage{graphicx}	
\usepackage{amsmath}	
\usepackage{longtable}

\title[]{The Type 1 and Type 2 AGN dichotomy according to their ZTF optical variability}
\author[E. López-Navas et al.]{
E. López-Navas,$^{1,2}$\thanks{E-mail: elena.lopez@postgrado.uv.cl}
P. Arévalo,$^{1,2}$
S. Bernal,$^{1}$
Matthew J. Graham,$^{3}$
L. Hernández-García,$^{4,1}$
P. Lira,$^{5,2}$
\newauthor
and P. Sánchez-Sáez.$^{6,4}$
\\
$^{1}$Instituto de Física y Astronomía, Facultad de Ciencias, Universidad de Valparaíso, Gran Bretaña 1111, Valparaíso, Chile\\
$^{2}$Millennium Nucleus on Transversal Research and Technology to Explore Supermassive Black Holes (TITANS)\\
$^{3}$California Institute of Technology, 1200 E. California Blvd, Pasadena, CA 91125, USA\\
$^{4}$Millennium Institute of Astrophysics (MAS), Nuncio Monseñor Sótero Sanz 100, Providencia, Santiago, Chile \\
$^{5}$Departamento Astronomía, Universidad de Chile, Casilla 36D, Santiago, Chile\\
$^{6}$European Southern Observatory, Karl-Schwarzschild-Strasse 2, 85748 Garching bei München, Germany\\
}

\date{Accepted XXX. Received YYY; in original form ZZZ}

\pubyear{2022}

\begin{document}
\label{firstpage}
\pagerange{\pageref{firstpage}--\pageref{lastpage}}
\maketitle

\begin{abstract}
The scarce optical variability studies in spectrally classified Type 2 active galactic nuclei (AGNs) have led to the discovery of anomalous objects that are incompatible with the simplest unified models (UM). This paper focuses on the exploration of different variability features that allows to separate between obscured, Type 2 AGNs, and the variable, unobscured Type 1s.  We analyse systematically the Zwicky Transient Facility, 2.5 years long light curves of $\sim$ 15000 AGNs from the Sloan Digital Sky Survey Data Release 16, which are generally considered Type 2s due to the absence of strong broad emission lines (BELs). Consistently with the expectations from the UM, the variability features are distributed differently for distinct populations, with spectrally classified weak Type 1s showing 1 order of magnitude larger variances than the Type 2s. We find that the parameters given by the damped random walk model leads to broader H$\alpha$ equivalent width for objects with $\tau _g$ > 16 d and long term structure function SF$_{\infty,g}$> 0.07 mag. By limiting the variability features, we find that $\sim$ 11 per cent of Type 2 sources show evidence for optical variations. A detailed spectral analysis of the most variable sources ($\sim$1 per cent of the Type 2 sample) leads to the discovery of misclassified Type 1s with weak BELs and changing-state candidates. This work presents one of the largest systematic investigations of Type 2 AGN optical variability to date, in preparation for future large photometric surveys.

\end{abstract}


%

\section{Introduction}

In recent years, the classical view of active galactic nuclei (AGNs) defined by the Unified Model \citep[UM,][]{antonucci1993} has been challenged by the discovery of various anomalous objects. 
Within the framework of the UM, an anisotropic, geometrically and optically thick dusty torus can hide, partially or completely depending on the line--of--sight of the system, the direct emission from the accretion disc and the broad line region (BLR). Phenomenologically, this allows to classify AGNs depending on their optical/\textit{UV} properties: sources with a blue continuum and broad emission lines (BELs) are called Type 1, whereas obscured, Type 2 AGNs are characterized by the lack of BELs in their optical/\textit{UV} spectra. 

The dusty torus obscuration corresponds to a typical extinction of 5--mag in the V band, and to an equivalent absorbing column density of NH$>10^{22}$ cm$^{-2}$ in the X--ray band for typical dust-to-gas ratios \citep{1995predehl}. In general, there is a good agreement between the X--ray absorption and the Type 1 and Type 2 dichotomy \citep[94 per cent of agreement for the majority of Seyferts in the \textit{Swift}--BAT AGN Spectroscopic Survey,][]{2017koss}. However, there is a few percent of X--ray unobscured AGNs whose optical spectra resemble Type 2 AGNs \citep{2001Pappa, 2002xia,2005wolter,2009panessa,2008brightman, 2008bianchi}. For most of these cases, the BELs are weak and are diluted by host galaxy contamination or by low signal to noise spectra \citep{2010Shi, barth2014}, so these sources are misclassified Type 1 AGNs. Interestingly, there are still a few more exotic sources, called \textit{naked} or \textit{true Type 2} AGNs, that \emph{could} intrinsically lack the denser gas that gives rise to the BLR, and/or the photoionizing continuum radiation that drives the broadline emission \citep{2010Shi,2011tran}. 


Independently of the origin of these unusual X--ray unobscured Type 2 AGNs, the discovery of the so-called changing-look (CL)/changing-state (CS) AGNs has evidenced that several scenarios can lead to variations in the BELs and so to the AGN optical classifications. In principle, the changes in the BELs could be associated with  a  large  change  of transient dust obscuration along the line-of-sight \citep{2019Yang, 2019wang}, in a similar way to the observed for the CL AGNs in X--ray astronomy \citep[e.g.][]{2015rivers}. Transient events such as tidal disruption events (TDEs) of a star by the supermassive black hole (SMBH) have also been claimed as possible drivers of CL phenomena  \citep[e.g. the case of 1ES 1927+654, which was previously classified as \textit{true Type 2}, ][]{trakhtenbrot2019}. However, the studies of most other optical CL AGNs have ruled out these scenarios in favor of intrinsic changes to the accretion flow. In this case, variations in the accretion rate would lead to a disappearing BLR or to a dimming (brightening) of the AGN continuum and so to a reduced (increased) supply of ionizing photons available to excite the gas around the SMBH \citep{ lamassa2015,runnoe2016,macleod2016,2017sheng,hutsemekers2017,Noda2018,hutsemekers2019,graham2020}. As mentioned above, this scenario also explains the existence of  \textit{true Type 2} AGNs. For instance, \citet{2021guolo} found that NGC 2992 transitions recurrently from Type 2 to intermediate-type at an Eddington ratio  ($\lambda$Edd) of $\sim$1 per cent. This means that at lower values of $\lambda$Edd, the AGN is still unobscured but intrinsically lacks BELs, which is by definition a \textit{true Type 2} AGN.


Another hallmark in AGNs that is heavily affected by the obscuration of the system is the observed temporal variability. According to the UM, the continuum coming from the central source in obscured AGNs is blocked by the dusty torus, so the optical variability is highly suppressed in Type 2 sources \citep[this prediction has been confirmed observationally, ][]{2009yip, 2017sanchez}. Based on these considerations, optical variability in spectroscopic Type 2 AGNs has led to the finding of sources that in principle could challenge the UM such as \textit{true Type 2} candidates \citep{2004hawkins}, misclassified Type 1 AGNs \citep{barth2014} and CL/CS AGNs \citep{lopez2022}. 

In this paper, we take advantage of the real time, deep, large sky-coverage monitoring survey Zwicky Transient Facility \citep[ZTF,][]{bellm2014,bellm2019}, to carry out one of the largest systematic investigations of Type 2 AGN optical variability to date, in comparison to a weak-Type 1 sample. The ZTF had first light in 2017 and employs an extremely wide $\sim$ 47  deg$^2$ field-of-view camera mounted on the Samuel Oschin 48-inch Schmidt telescope. It is designed to scan the entire Northern sky every two days in the \textit{gri} filters, which  enables  a  wide  variety  of  novel  multiband  time-domain studies. Here, we investigate different variability features that help us distinguish between Type 1 and Type 2 AGNs with the ZTF light curves, which leads to the discovery of a new sample of weak Type 1 AGNs and CL/CS candidates.



\section{Sample and data}\label{sec:sample} 


The parent sample consists of the 30520 galaxies classified as GALAXY AGN in the Sloan Digital Sky Survey (SDSS) Data Release 16 \citep[DR16,][]{DR16}, which have detectable emission lines that are consistent with being a Seyfert or LINER according to the BPT-type \citep{1981bpt} criteria employed by the SDSS pipeline  (log$_{10}$(OIII/H$\alpha$)$ <$ 0.7 – 1.2(log$_{10}$(NII/H$\alpha$) + 0.4). 
These sources were not identified with a quasar (QSO) template by the SDSS pipeline, so they are generally weaker or obscured AGNs \citep[we refer to ][for details on how the pipeline classifies each spectrum]{2012bolton}. Light curves for the 29057 sources observed by the ZTF in the \textit{g} and \textit{r} filters were extracted from a forced photometry data set that has been produced based on all difference images available for ZTF DR5 (Mroz et al., in prep) and source positions from the Panoramic Survey Telescope and Rapid Response System (Pan-STARRS)  Data Release 1 catalog \citep[PS1 DR1][]{PS1}. To clean the light curves for bad photometry data points we performed a 3-$\sigma$ clipping twice and discarded the points whose error was greater than twice the mean value of the errors from each particular light curve. The variability analysis was performed on the 3-days binned light curves to reduce the errors on the light curves data points. After the cleaning and binning, we discarded the light curves with <10 data points, which reduces the sample to 20583. The final light curves have a mean of 100 data points in each band, spanning 900 days between April, 2018 and December, 2020.

In the sample, 2737 sources are sub-classified as BROADLINE by the SDSS pipeline, which means that lines with width $> 200$ km s$^{-1}$  can be detected at least at the 5-sigma level. Given the relatively low threshold for the line widths, these do not necessarily correspond to Type 1 AGNs. Additionally, broad lines detected at lower significance are not reported, so the pipeline cannot identify \emph{weak} Type 1 AGNs. Therefore, more detailed analyses are needed to properly separate the different optical types of AGNs in the SDSS sample. 

Here, we differentiate various subsamples according to previous studies and catalogues. In particular, we cross-matched the sample within a 1 arcsec radius with the Type 1s with weak BLRs from \citet{Oh2015} and \citet{2019liu}, selected via detailed modelling of SDSS DR7 spectra. We note that most of the Type 1 AGNs from \citet{Oh2015} were also in \citet{2019liu}'s catalogue, so we merged both subsamples together (the \textit{Weak Type 1} sample). These studies are based on the existence of a significant broad component and not to the width of the BEL, leading to the discovery of Type 1s at the low-mass and low-luminosity end. In particular, \citet{2019liu} report H$\alpha$ luminosities in the range 10$^{38.5}$–10$^{44.3}$erg s$^{-1}$, and line widths (FWHMs) of 500–34,000 km s$^{-1}$. In spite of the wide FWHMs and similar requirement on line significance, only 20 per cent of sources in the Weak Type 1 sample were classified as BROADLINE in the SDSS. On the other hand, 50 per cent of BROADLINE galaxies have not been identified as Type 1 in these catalogues despite they were presumably in their parent sample (galaxies and QSOs from the SDSS DR7, October 2008). Then, the difference in the selection of weak Type 1 and BROADLINE sources is not clear, and it might just arise from a more detailed spectroscopic method, with a better substraction of the galaxy contribution by \citet{Oh2015} and \citet{2019liu}.

\begin{figure}
 \includegraphics[width=\columnwidth]{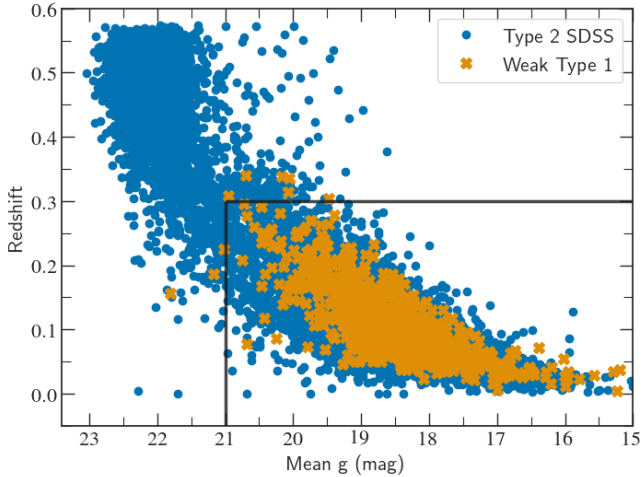}
 \caption{Redshift versus mean \textit{g} magnitude for different samples considered in this work. The solid black lines indicate the limits used to compare different samples, z<0.3 and Mean \textit{g} < 21 mag.}
 \label{fig: zmag}
\end{figure}

We also separated the sources subclassified as BROADLINE in the SDSS and the blazars from The Roma-BZCAT Multi-Frequency Catalog of Blazars \citep[\textit{Blazar ROMA},][]{2015massaro}. In Table~\ref{tab: samples} we show the samples considered in this work, with the one called \textit{Type 2 SDSS} being the parent sample without the sources that are in the other subsamples (BROADLINE, Blazar ROMA, Weak Type 1 and AGN lcc, the latter being explained in the next section). We note that, while the Weak Type 1 sample is quite pure, a level of contamination of very low-luminosity Type 1 objects is expected in the Type 2 sample. Since this work is focused on the comparison between the Type 1 and Type 2 objects, we selected the mean magnitudes and redshifts to be comparable between the samples: $g<21$ mag, $r<20$ mag and $z<0.3$ (see Figure~\ref{fig: zmag}).
The final number of sources analysed after this selection are listed in Table~\ref{tab: samples}. We note that, except for the Type 2 SDSS sources, some AGNs are included in more than one subsample. 

%
 

%
\begin{table}
	\centering
	    \begin{adjustbox}{max width=\columnwidth}
	    \begin{threeparttable}
	\caption{Samples analysed in this work.}
	\label{tab: samples}

	\begin{tabular}{ccc} 
			\hline
		Name& N$^\circ$ of sources& g<21 / r<20 mag\\
		&  &  $\&$ z<0.3	\\ 
		\hline
		Type 2 SDSS &16999 &12743 / 13322 \\
		Weak Type 1 & 978 & 921 / 960 \\
		Blazar ROMA &26 &20 / 21 \\
		BROADLINE & 2737 & 2014 / 2118  \\
		AGN lcc & 260 & 245 / 251 \\
	\end{tabular}
	\end{threeparttable}
	\end{adjustbox}
\end{table}

\section{Light curve variability}
We performed the variability analysis of the ZTF light curves on the SDSS DR16 AGN sample by extracting some of the variability features used by the Automatic Learning for the Rapid Classification of Events \citep[ALeRCE,][]{forster2021} broker light curve classifier \citep[LCC, ][]{sanchez2021}. The ALeRCE broker is currently processing the alert stream from the ZTF and has been selected as a Community Broker for the Vera C. Rubin Observatory and its Legacy Survey of Space and Time (LSST). The goal of the LCC is to provide a fast classification of transient and variable objects by applying a balanced random forest algorithm. For this purpose, a total of 174 features, including variability features and  colors obtained from AllWISE and ZTF photometry, are computed for every object with at least 6 alerts in either \textit{g} or \textit{r} band. The complete set of features are described in \url{http://alerce.science/features/}, and a python library to extract variability features in astronomical light curves is publicly available at  \url{https://github.com/alercebroker/lc_classifier}. In this work, we selected a set of variability features included in the LCC python library and applied them on the 3-days binned \textit{g} and \textit{r} ZTF light curves. For comparison, we also cross-matched our DR16 AGN sample with the sources classified as AGN or Blazar by the LCC, and treat them as another subsample in our variability analysis (\textit{AGN lcc}). We note that this is the only subsample selected by flux variability, as opposed to spectral features, considered in this work.

\subsection{Variability features}
We characterize the optical variability of the ZTF light curves by extracting the following features included in the LCC python library:

\begin{itemize}
    \item Excess Variance (ExcessVar): measure of the intrinsic variability amplitude. In the LCC library it is defined as ExcessVar$_{LCC}=({\sigma }_{\mathrm{LC}}^{2}-{\overline{\sigma }}_{m}^{2})/{\overline{m}}^{2}$, where $\sigma _{\mathrm{LC}}$ is the standard deviation of the light curve, ${\overline{\sigma }}_{m}$ is the average error, and $\overline{m}$ is the average magnitude. Here, we use  ExcessVar$=({\sigma }_{\mathrm{LC}}^{2}-{\overline{\sigma }}_{m}^{2})$, which is a magnitude version of excess variance analogous to the normalized excess variance that is typically used in linear flux units \citep{2017sanchez}. Similar features have been broadly used to study X--ray variability \citep[and references therein]{1997nandra, 2002edelson,2003vaughan}.
    \item Damped random walk (DRW) parameters: the DRW model is generally used to describe the AGN optical variability \citep[see][and references therein]{2009kelly}. The model considers the light curves as continuous time stochastic processes, and provides an estimation of the characteristic timescale $\tau$ and amplitude square $\sigma^2$ of the variations. Here, we will report the asymptotic value of the structure function on long timescales (SF$_\infty$) given by SF$_\infty = \sqrt{2} \cdot \sigma$, which is broadly used in the literature \citep{2010macleod}.
    We also correct $\tau$ by the redshift of the sources as reported in the DR16 to obtain the intrinsic timescales of the variations in the rest frame.
    
    \item Mexican hat power spectrum \citep[MHPS,][]{Arevalo2012}: This method isolates variability on different timescales in unevenly sampled light curves by applying a Mexican-hat type filter. It convolves each light curve with two Gaussian profiles of slightly different widths and takes the difference of the convolved light curves. This difference is dominated by fluctuations at timescales of $\sigma$ /0.225, where $\sigma$ is the average width of the Gaussian filters, removing variations on shorter and longer timescales. The variance of the resulting difference, as a function of frequency, is an estimation of the power spectrum. The timescales by default are 10 and 100 days, which are convenient to separate AGNs to other stochastic sources such as long-period variable stars and young stellar objects. Since we will analyse just AGNs, we use the same feature than the LCC but we compute it at 300 (MH $\sigma^2_{300}$) and 150 days (MH $\sigma^2_{150}$).

\end{itemize}

\section{Results}

\begin{figure}
 \includegraphics[width=\columnwidth]{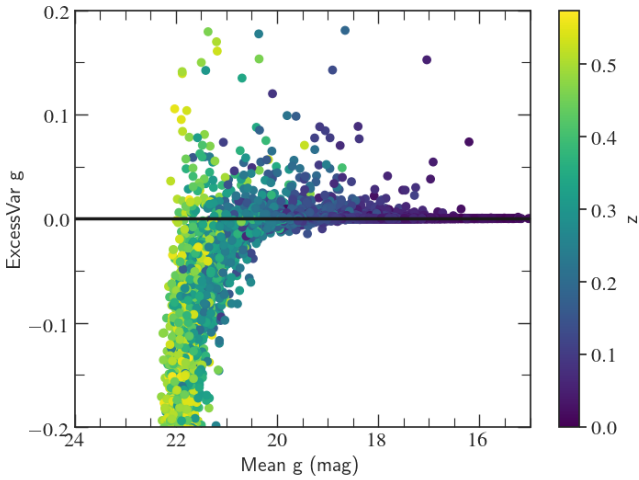}
 \caption{Excess variances (ExcessVar) versus mean magnitude in the \textit{g} band for the SDSS DR16 AGN sample ZTF light curves. Above 21 magnitudes, most of the excess variances are negative, which indicates an overestimation of the average errors.}
 \label{fig:excess_mag}
\end{figure}

In order to characterize the optical variability of the SDSS DR16 AGN sample, we first computed the Excess Var for all the objects. This feature evaluates the variance after subtracting the contribution expected from measurement errors. As it can be seen in Figure~\ref{fig:excess_mag}, the excess variance distributes approximately symmetrically around zero for magnitudes below \textit{g} $\sim$21 mag, with a few sources showing clear positive values.
This is expected when the intrinsic variance is essentially zero, so this behaviour is produced by an unbiased uncertainty in the noise estimate. In contrast, at dimmer magnitudes the Excess Var drops toward negative values, showing that the noise is systematically overestimated at larger values than the ZTF limiting magnitude \citep[g $\sim$20.8 mag, AB, 5 $\sigma$ in 30 sec,][]{masci2019} . 

After the redshift and magnitude selection, we computed the variability features for each sample, whose results are presented in Table~\ref{tab: results}. As it can be seen, the samples are significantly distinct, with one order of magnitude difference between the variances of the less (Type 2 SDSS) and the most (AGN lcc) variable sources. Notably, the BROADLINE and the Type 2 SDSS samples present very similar features, with lower variances and DRW parameters than the Weak Type 1 objects. In parallel, the features from the Blazar ROMA sample are comparable to those of the Weak Type 1s.

Since the Weak Type 1 sample contains the only objects that have been classified via detailed spectroscopy, we next focus on the differences between this and the Type 2 SDSS sample to examine the ability of the selected features to separate both populations.

\begin{table*}
	\centering
	    \begin{adjustbox}{max width=0.97\textwidth}
	    \begin{threeparttable}
	\caption{Variability results for each sample considered in this work. The light curves were selected to have <21 mag in the \textit{g} band, <20 mag in the \textit{r} band, and z<0.3. The values reported correspond to the 16th, 50th and 84th percentiles. }
	\label{tab: results}

	\begin{tabular}{ccccccccccc} 
			\hline
		Sample&Counts&ExcessVar g&ExcessVar r& MH $\sigma^2_{g300}$& MH $\sigma^2_{r300}$& SF$_{\infty,g}$&SF$_{\infty,r}$& $\tau_g$& $\tau_r$ \\
		& g/r & $\cdot 10^{-5}$ &$\cdot 10^{-5}$ &$\cdot 10^{-2}$ &$\cdot 10^{-2}$& mag &mag & days & days	\\ 
		\hline
	     \textbf{Type 2 SDSS}&12743/13322&$7^{60}_{-40}$&$7^{24}_{-0.7}$& $0.17^{1.6}_{-0.5}$&$0.10^{0.5}_{-0.07}$&$0.018^{0.04}_{1.9e-6}$&$0.008^{0.018}_{4e-8}$&$6^{50}_{1.1}$&$5^{30}_{1.2}$ \\
		\hline
		
		\textbf{BROADLINE}& 2014/2118 &$10^{70}_{-25}$&$8^{30}_{0.9}$& $0.16^{3}_{-3}$&$0.10^{1.0}_{-0.21}$&$0.018^{0.05}_{4e-6}$&$0.010^{0.021}_{9e-7}$&$7^{60}_{1.3}$&$5^{30}_{1.2}$\\
		\hline
		\textbf{Weak Type 1}&921/960&$60^{500}_{1.2}$&$30^{140}_{5}$& $1.3^{14}_{0.012}$&$0.5^{3}_{0.03}$&$0.04^{0.11}_{4e-5}$&$0.019^{0.05}_{7e-7}$&$30^{100}_{3}$&$21^{100}_{2.4}$\\
		\hline
		\textbf{Blazar ROMA}&20/21$70^{300}_{40}$&$37^{300}_{15}$ & $1.5^{8}_{0.7}$&$0.7^{5}_{0.15}$&$0.06^{0.08}_{0.03}$&$0.03^{0.09}_{0.010}$&$15^{50}_{7}$&$30^{80}_{7}$&\\

		\hline
		\textbf{AGN lcc}&245/251&$500^{2200}_{90}$&$170^{600}_{40}$&$12^{70}_{1.2}$ & $3^{20}_{0.7}$&$0.11^{0.24 	}_{0.04}$&$0.06^{0.12}_{0.024}$&$70^{180}_{19}$&$60^{180}_{19}$\\
		\hline

	\end{tabular}
    
       \end{threeparttable}
       \end{adjustbox}

\end{table*}

\subsection{Type 2 versus Weak Type 1 variability features}
\begin{figure*}
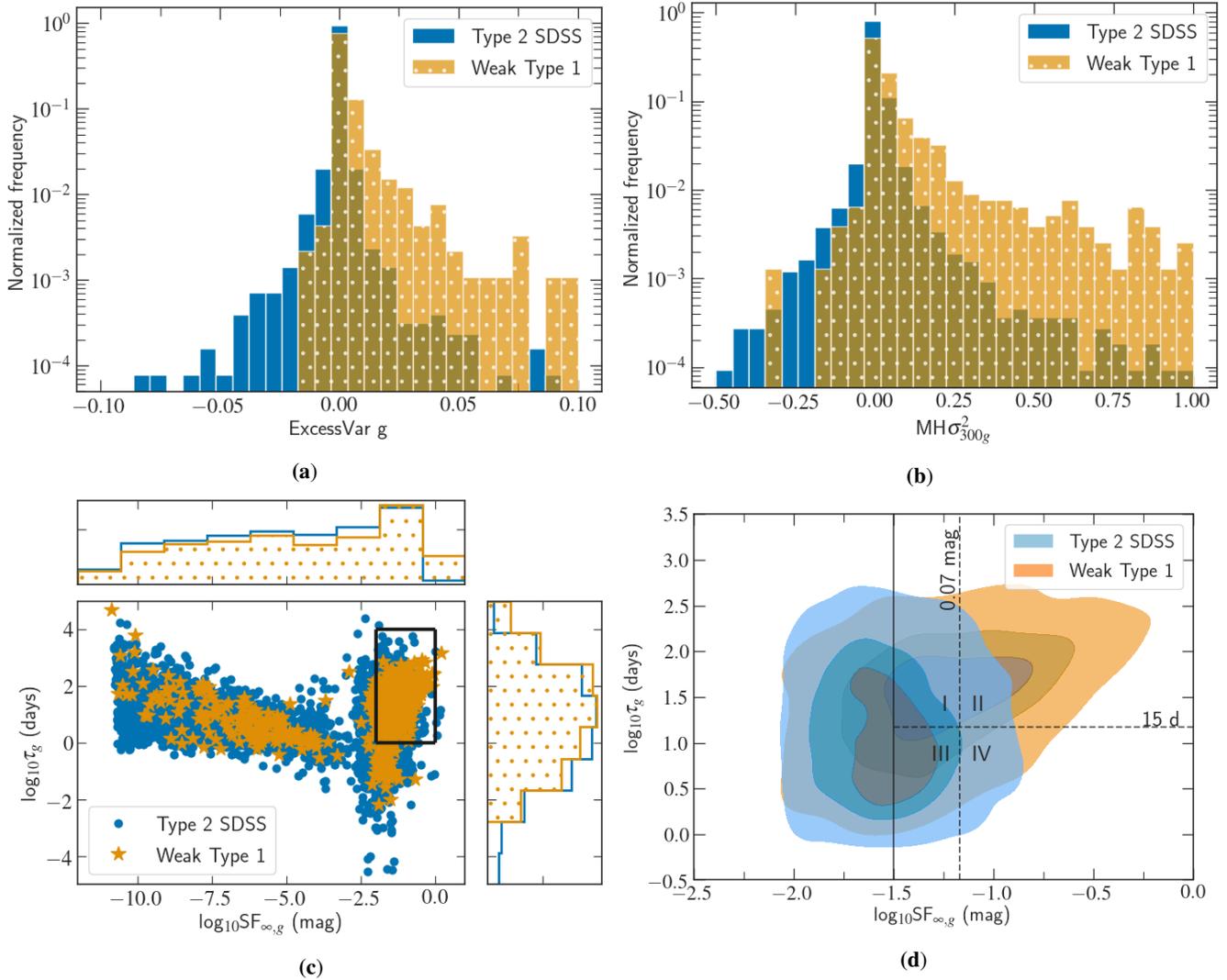

        \centering
        \begin{subfigure}{0.49\textwidth} 
            \includegraphics[width=\textwidth]{plots/hist_excessvarg.pdf}
            \caption{\normalsize{\textbf{(a})}}
            \label{fig:var_dampa}
        \end{subfigure}       
        \begin{subfigure}{0.49\textwidth} 
            \includegraphics[width=\textwidth]{plots/hist_MH300_t1t2_g.pdf}
            \caption{\normalsize{\textbf{(b})}}
            \label{fig:var_dampb}
        \end{subfigure}
                \begin{subfigure}{0.48\textwidth} 
            \includegraphics[width=\textwidth]{plots/sfinf_all_hist.pdf}
            \caption{\normalsize{\textbf{(c})}}
            \label{fig:var_dampc}
        \end{subfigure}       
        \begin{subfigure}{0.48\textwidth} 
            \includegraphics[width=\textwidth]{plots/DRWg_contours_tauz.pdf}
            \caption{\normalsize{\textbf{(d})}}
            \label{fig:var_dampd}
        \end{subfigure}

        \caption{Comparison of the variability features in the \textit{g} band for the Type 2 SDSS and the Weak Type 1 samples. (a) Excess variance (b) Variance at the 300 days timescale. (c) Damped random walk (DRW) parameters:  intrinsic timescale ($\tau_g$) of the variations versus the long term structure function (SF$_{\infty,g}$). The black square indicates the region enlarged in sub-figure (d). Histograms at the sides of the plot show the normalized frequency of each parameter. (d) DRW parameters within the limits $0.01 <$ SF$_{\infty,g} < 1$ mag and $ 1< \tau_g < 10^{4}$ d. The solid black line indicates the DRW region chosen to investigate the spectral properties of the Type 2 SDSS sample ($0.03 <$ SF$_{\infty,g} < 1$ mag), and the dashed lines show the different subdivisions used in the spectral analysis. The distinct behaviour of these variability features reinforces both samples belong to different populations.  } 
    \end{figure*}

Firstly, we investigated the Excess Var for the two samples, which is shown in Figure~\ref{fig:var_dampa}. From now on, we will present the results just in the \textit{g} band, but the same conclusions are recovered when studying the \textit{r} band. In this figure, it can be seen that for the Type 2 SDSS sample a small fraction of sources with large negative excess variances remains, but in general the data points distribute around zero as expected for non-variable sources.
Although both samples have objects with positive and negative excess variances, the distribution of Weak Type 1s is significantly skewed towards positive values, indicating that as a sample they have stronger variability. A Kolmogorov-Smirnov (ks) test demonstrates the significance of this difference with ks=0.35 and p-value=2$\cdot10^{-93}$. We tested whether the difference in the samples could be due to the different distributions in magnitude, in particular to the higher fraction of Type 2 sources in the range 20<g<21 mag, where the Excess Var starts to drop to negative values. Cutting both samples at g<20 and repeating the ks test resulted in ks=0.34 and p-value=3$\cdot10^{-88}$, showing that the lower magnitudes in the Type 2 sample are not producing the difference in the Excess Var. Similar results are obtained by analyzing the MHPS variances computed at the intrinsic timescales of 300 and 150 days (see Figure~\ref{fig:var_dampb} for the 300 d timescale). For the 300 (150) d timescale in the \textit{g} band, the ks test results in ks=0.33 (0.32) and p-value=4$\cdot10^{-69}$ (5$\cdot10^{-65}$), which demonstrates again the significance of the distinction between the samples.


Finally, we examined the characteristic timescale $\tau$ and the long term structure function (SF$_{\infty}$) of the variations, given by the DRW model. In Figure~\ref{fig:var_dampc} we show the DRW parameters in the \textit{g} band for the two samples. It can be seen that many points take non-physical values, with extremely small SF$_{\infty}$ (for example all the cloud of points with SF$_{\infty}< 10^{-2.5}$ mag ) or timescales smaller than the actual data sampling (3 days bins) or larger than the maximum light curve length (1000 days). These results indicate that the DRW model is fitting noise primarily for most of the Type 2 SDSS sources, and not real variations. Moreover, the normalized histograms show a very similar behaviour for both samples except for the limit of highest SF$_{\infty}$, which is significantly more populated in the Weak Type 1 sample.

To investigate whether any reliable features can be recovered from these results, we compared the DRW parameters of the Type 2 SDSS and Weak Type 1 samples within the broad but more physically meaningful limits of $0.01 <$ SF$_{\infty,g} < 1$ mag and $ 1< \tau_g < 10000$ d (black square in Figure~\ref{fig:var_dampc}). The outcome of this exercise is plotted in Figure~\ref{fig:var_dampd}. In spite of the overlap between the samples, the distribution of Weak Type 1s shifts to higher SF$_{\infty,g}$ and $\tau_g$ values than the Type 2 SDSS sample, which shows that AGNs with weak BELs in their spectra correspond to a more variable population in flux. 

These results confirm the expectation that the weak BELs indeed correspond to a Type 1 activity where the variable continuum emission is also visible. Additionally, weak Type 1 AGNs can be identified through their optical light curves, allowing in principle the selection of weakly accreting black holes from existing and future large photometric variability surveys.

\subsection{Subsamples selected through variability}
\subsubsection{DRW parameters and spectral properties}\label{drw}
The comparison between the Type 2 SDSS and the Weak Type 1 samples suggests that both classes are distributed differently in the DRW space. To evaluate this result, we can model the optical spectra of two different groups selected by their DRW parameters and determine whether their variability and spectral properties correlate. 


To this end, we selected the Type 2 SDSS sources that overlap in the DRW space with the locus of the Weak Type 1 distribution at log$_{10}$SF$_{\infty,g} > -1.5$ (black solid line in Figure~\ref{fig:var_dampd}) with well determined \textit{g} and \textit{r} DRW parameters as follows: \begin{itemize}
    \item $g<20$ mag, to select the brightest sources,
    \item   $-1.5 < $ log$_{10}$ SF$_{\infty,g,r} < 0$ (most variable sources) and 
    \item $ 0< $ log$_{10}$ $\tau_{g,r} < 3$ (significant timescales).
\end{itemize}
 

We obtained 353 sources that met these requirements, which comprises 2 per cent of Type 2 SDSS sources (353/16999). In comparison, there are 28 per cent (272/978) of Weak Type 1s within this DRW space.  

From the 353 Type 2s, we selected randomly 160 objects to perform the spectral analysis, 80 with $\tau _{g}< 15$ d  and 80 more with $\tau _{g}>15$ d, which corresponds to the value that separates most of the Weak Type 1 sources as shown in Figure~\ref{fig:var_dampd}. We downloaded the archival SDSS spectra available for each source and fitted the spectra using the Penalized Pixel-Fitting (pPXF) software \citep{cappellari2017improving}. To model the spectra we used the E-MILES library \citep{2016MNRAS.463.3409V} to account for the stellar continuum component, a set of power law templates for the accretion disc contribution and two components for the emission lines, one with both permitted and forbidden lines to model the narrow emission and one just with the permitted lines to model the possible BELs. To compute the fitting errors, a total of 50 Monte Carlo simulations were performed for each spectrum using the residual of the best-fit to generate random noise. This noise was then added to the original spectrum and fitted with the same procedure. 

\begin{figure*}
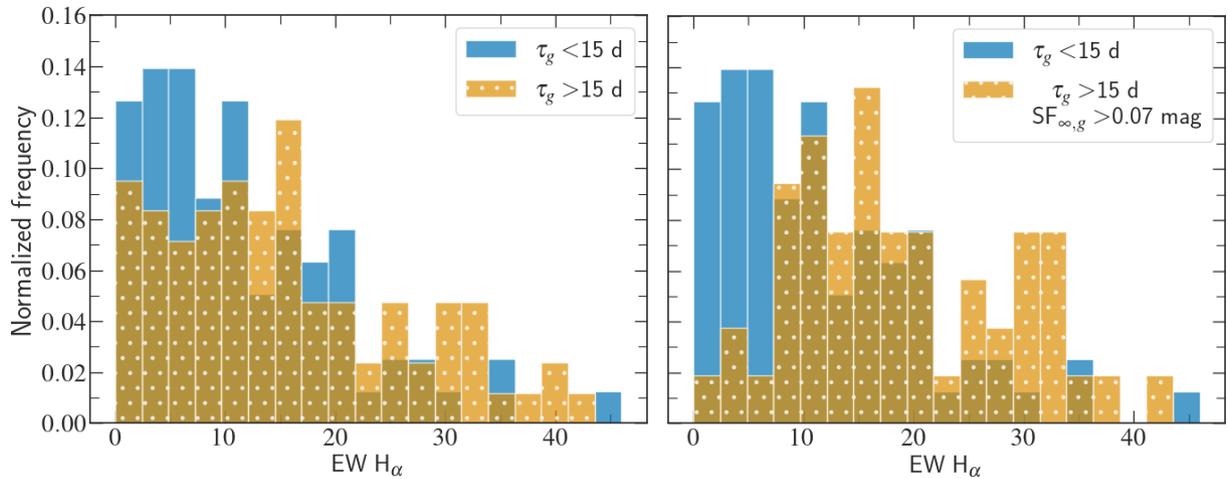

 \includegraphics[width=\columnwidth]{plots/tha_hist.pdf}
  \includegraphics[width=0.88\columnwidth]{plots/stha_hist.pdf}
 \caption{Normalized histograms for the H$\alpha$ equivalent width of two different subsamples of the Type 2 SDSS sample selected according to their DRW parameters. In both cases, we recover different distributions, with the EW H$\alpha$ of the sources outside the Weak Type 1 locus in the DRW space peaking at lower values than the sample that overlaps with the Type 1s  ($\tau _{g}> 15$ d and SF$_{\infty, g}> 0.07$ mag).  }
 \label{fig:ha_hist}
\end{figure*}

    In Figure~\ref{fig:ha_hist} we show the distribution of the broad H$\alpha$ equivalent width (EW H$\alpha$) for the Type 2 subsamples separating them at $\tau _{g}= 15$ d (region I+II vs III+IV), and at the region delimited by both $\tau _{g}= 15$ d and SF$_{\infty, g}= 0.07$ mag (region II vs III+IV). As supported by the ks test shown in Table~\ref{tab: ks}, the distribution of EW H$\alpha$ is consistently different between the two samples. In both cases, the spectroscopic analysis leads to a EW H$\alpha$ distribution skewed towards larger values for the sources whose DRW parameters overlap those of the Weak Type 1s, with a higher significance when applying limits to both $\tau _{g}$ and SF$_{\infty, g}$ simultaneously (i.e., region II vs III+IV). We also evaluated the existence of a relationship between the individual EW H$\alpha$ and the DRW parameters themselves, but no clear correlation was found.

\begin{table}
	\centering
	    \begin{adjustbox}{max width=\columnwidth}
	    \begin{threeparttable}
	\caption{Kolmogorov-Smirnov (ks) test for different groups of the Type 2 SDSS sample selected according to their DRW parameters.}
	\label{tab: ks}

	\begin{tabular}{cccccc} 
			\hline
		\textbf{Regions} & I+II & II+ IV& II & II \\
		& III+IV& I+III& III+IV &  III  \\
		\hline
		\textbf{N$^\circ$ sources}& 80/80 & 97/63 & 50/80 & 50/33  \\
		\textbf{ks} & 0.20 & 0.35 & 0.36 & 0.47 \\
		\textbf{p-value} & 0.04& 0.0001 & 0.0003 & 0.0001\\
	\end{tabular}
	\end{threeparttable}
	\end{adjustbox}
\end{table}

\subsubsection{Most variable sources}
\begin{figure*}
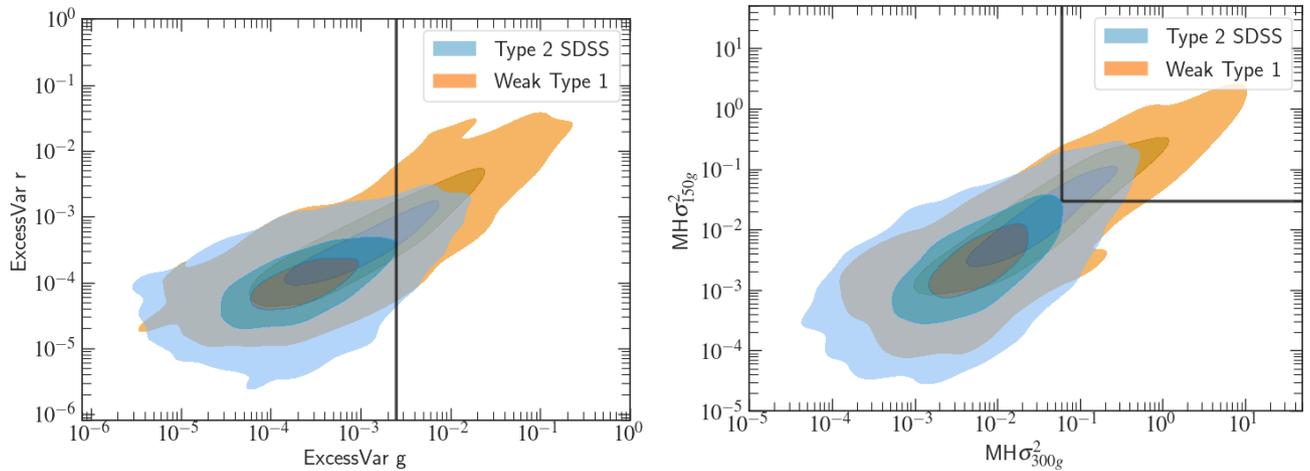


  \includegraphics[width=\columnwidth]{plots/excessvar_contours_new.pdf}
  \includegraphics[width=\columnwidth]{plots/MH_t1t2g.pdf}

 \caption{Comparison of the positive variances for two different samples of AGNs. Left: Excess variances in the \textit{r} band against the \textit{g} band. Right: \textit{g} band variances at 150 d timescale versus at 300 d timescale. It can be seen that all the parameters distribute around higher values for the Weak Type 1 sample. The black solid lines indicate the limiting values of variances in the \textit{g} band to select the most variable AGNs, which exclude the 67 per cent limit of Type 2 sources with positive variances.}
 
 \label{fig:variances}
\end{figure*}

The variability features used in this work have been shown to be a very powerful tool to separate different AGN populations. As a last step of the analysis, we selected the most variable sources of the entire sample to investigate their nature. We searched for current optical variability of sources that previously looked as Type 2 due to the absence of significant BELs in their spectrum, so they could be Type 1 AGNs that were misclassified, or CL/CS AGNs that have changed their type since their SDSS spectrum was taken. To determine the criteria to select the most variable sources, we first compared the positive variances for the Type 2 SDSS and Weak Type 1 samples, as illustrated in Figure~\ref{fig:variances}.  Although we find an overlap between the distributions, we recover higher ExcessVar$_{g,r}$ and MH $\sigma^2_{300,150g}$ for the Weak Type 1 sample. Here, we take as the limiting value for each variance the 2/3 level of the Type 2 SDSS sample, that is, which excludes the 67 per cent of Type 2 sources with positive variances. As shown in Figure~\ref{fig:variances}, the limiting values are found to be ExcessVar $g>2.5\cdot 10^{-3}$ , MH $\sigma^2_{300g} >6\times 10^{-2} $ and MH $\sigma^2_{150g} >3\times 10^{-2} $. 
Then, we applied the following criteria simultaneously to select the most variable sources (in the \textit{g} band):
\begin{itemize}
    \item Mean $g <21$ and $z<0.3$
        \item ExcessVar $g>2.5\cdot 10^{-3}$ 
    \item $0.07 < $ SF$_{\infty,g} < 1$ mag and $ 15< \tau _g < 1000$ d 
    \item MH $\sigma^2_{300g} >6\cdot 10^{-2} $ and MH $\sigma^2_{150g} >3\cdot 10^{-2}$
\end{itemize}

\begin{figure*}
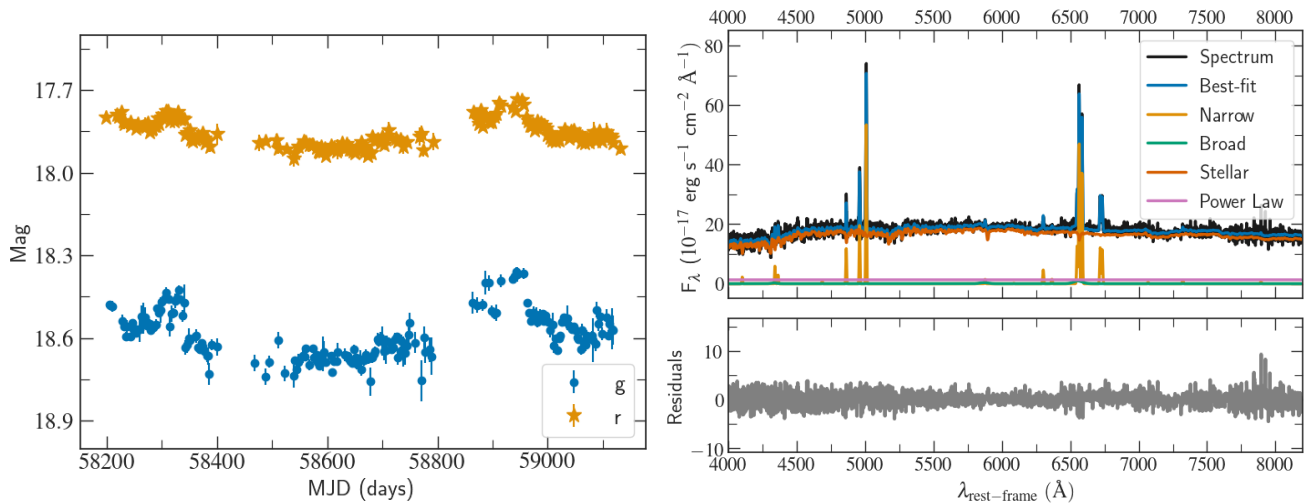


  \includegraphics[width=\columnwidth]{plots/lc1.pdf}
  \includegraphics[width=\columnwidth]{plots/spectrum1.pdf}

 \caption{ZTF light curves (left) and fit to the archival SDSS spectrum (right, MJD=52443) of one of the CL candidates found in this work, J161219.56+462942.62. The light curves show significant aperiodic variability typical of Type 1 AGNs, while the spectrum shows very weak broad emission lines.}
 
 \label{fig:cl}
\end{figure*}

We obtained 246 sources that met the requirements. From these, there are 133 that have been classified as Type 1 in \citet{Oh2015} and/or in \citet{2019liu}, 10 more that have been subclassified as BROADLINE in the SDSS and 2 as blazar in Roma-BZCAT. From the remaining 98 sources, we discarded 12 that were classified as Sy1, QSO, LINER or BLLac in the SIMBAD database. We also checked that none of the 89 remaining sources were classified as QSO (type-1 broad-line core-dominated) or AGN (type-1 Seyferts/host-dominated) in The Million Quasars (MILLIQUAS) Catalog \citep{2021flesch}, Version 7.5 (30 April 2022). We inspected visually the light curves of the 89 sources and we performed a spectral analysis of their archival SDSS spectra, in a similar way as described above (see Section~\ref{drw}). This analysis led to a list of 77 most likely misclassified Type 1 AGNs, with EW H$\alpha$ > 5\AA ~(to be consistent with \citealt{2019liu}'s classification), and 
12 CL candidates, with very weak or absent BELs (EW H$\alpha$ 5< \AA). 
As an example, in Figure~\ref{fig:cl} we show the ZTF light curves and the fits to the SDSS spectrum for one of the CL candidates. The fitting results for all the sources and a brief discussion of some anomalous AGNs are presented in the Appendix \ref{appendix}. We note, however, that some of these sources could have been already reclassified in other works that are not cited in this paper.


\section{Discussion}
According to the simplest versions of unified models for AGNs, it is expected that optical variability in Type 2s should be suppressed due to the obscuration of the variable nuclear continuum by the dusty torus. In agreement with this picture, our results indicate that the Type 2 SDSS sample has much lower variances (given by the Excess Var and MH $\sigma^2_{300,150}$) and DRW parameters (the structure function on long timescales SF$_\infty$ and the timescale $\tau$ of the variations) than the Weak Type 1 and Blazar ROMA samples that overlap with the SDSS AGN (not QSO) classification. These results are also broadly consistent with the conclusions from \citet{2009yip}, who found no evidence of continuum or emission-line variability in Type 2 AGNs on timescales of months to a few years.
Specifically, the Type 2 sources present excess variances that are fairly symmetrical around zero, which is expected for non-variable objects: ExcessVar$_g = [-1000,-40,7,60,700]\cdot 10^{-5}$ for the [1, 16, 50, 84, 99]-th percentiles. Similarly, for the MHPS variances at 300 d timescale we get MH $\sigma^2_{g300} = [-12,-0.5,0.17,1.6,21]\cdot 10^{-2}$. In addition, the DRW parameters take values within a very broad range of amplitudes and timescales: SF$_\infty,g = [6\cdot 10^{-11}, 1.9\cdot 10^{-6}, 0.018, 0.04, 0.15] $ mag and $\tau_g = [0.06, 1.1, 6, 50, 800]$d. These extremely small values up to the 50th percentile imply that the model is fitting noise primarily in these non-variable sources, and just a fraction <50 per cent of  Type 2s have variations that can be characterized by physically meaningful values of amplitudes and timescales.

Moreover, we find that the BROADLINE objects have a very similar variability behaviour to the Type 2 SDSS, which suggests the BELs that the SDSS pipeline detects are not necessarily coming from the BLR and might correspond to outflows. These results are consistent with the fact that $\sim$ 30 per cent of Type 2s at z<1 identified by \citet{2016yuan} from the SDSS-III/Baryon Oscillation Spectroscopic Survey (BOSS) spectroscopic data base, selected on the basis of their emission-line properties, were subclassified as BROADLINE by the SDSS pipeline. We also considered to include \citet{2016yuan}'s spectroscopically confirmed Type 2s as another subsample in the analysis, but these sources were too dim (23 <g mag <21) to get reliable variability results with the ZTF data.  

On the other extreme of variability we have a subsample of objects that have been classified as AGN or Blazar according to the LCC (the AGN lcc sample). By construction, this sample is expected to be the most variable, since the ZTF produces alerts when a 5 $\sigma$ variation in the template-subtracted images is detected. Accordingly, the AGN lcc sources show the highest variances and DRW parameters of all, with values up to 2 orders of magnitude above those of the Type 2 SDSS sample. For the LCC training set of the AGN class, \citet{sanchez2021} considered the Type 1 Seyfert galaxies (i.e., AGNs whose emission is dominated by the host galaxy), selected from MILLIQUAS (broad type “A”), and from \citet{Oh2015}, and for the Blazar class they selected the BL Lac objects and Flat Spectrum Radio Quasars from The Roma-BZCAT Multi-Frequency Catalog of Blazars  and MILLIQUAS. The reported \textit{g}-band excess variances for both populations (Figure 20 in their paper) peak at $3 \cdot 10 ^{-5}$ (AGN) and  $5 \cdot 10 ^{-4}$ (Blazar), which are in agreement with our results for the AGN lcc sample: ExcessVar$_{LCC},g = 1.6 ^{6}_{0.3} \cdot 10^{-5}$ (for the 16, 50 and 84-th percentiles).

\begin{figure}
  \includegraphics[width=\columnwidth]{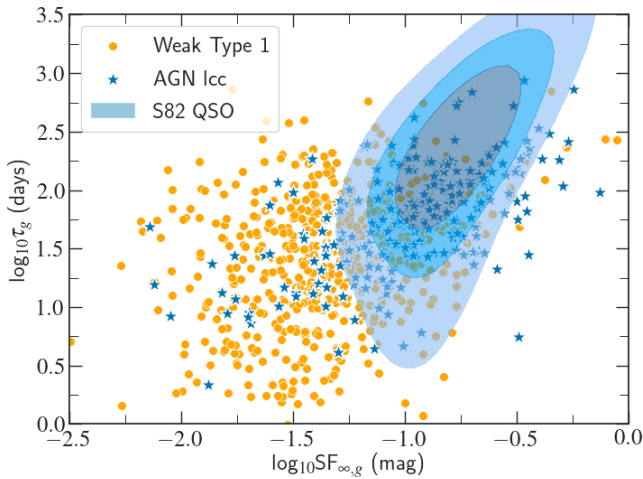}
 \caption{Comparison of the DRW parameters between different samples from this work and the SDSS Stripe 82 QSOs from \citet{2010macleod}}
 \label{fig:macleod}
\end{figure}

One of the most cited works that include the time variability analysis of AGNs via the DRW is that from \citet{2010macleod}, who model the optical variations of $\sim$9000 spectroscopically confirmed QSOs from the SDSS  Stripe  82 (S82). To this end, the authors analyse SDSS \textit{ugriz} photometric light curves with more than 60 epochs of observations over a decade. In Figure~\ref{fig:macleod} we show the comparison between the DRW parameters of the S82 QSOs, the Weak Type 1 and the AGN lcc samples, noting that this is an enlarged section of the entire DRW range. Since the S82 QSOs have a median redshift of 1.5, we selected the results for the infrared \textit{z} band (9134 \AA) to compare variations at similar emitted wavelengths to our z$<0.3$ sources in the \textit{g} band (4770 \AA).  As can be seen in the figure, the Weak Type 1s tend to have lower SF$_\infty$ and $\tau$ values than the S82 QSOs. We can also see that most of the sources of the AGN lcc sample overlaps with the S82 QSOs, and just a few sources lie at lower DRW parameters. This comparison highlights the differences between variations of Type 1 objects, where AGNs with weaker BELs have lower amplitudes and characteristic timescales than brighter QSOs. Specifically, just 16 per cent of Weak Type 1s varies strongly enough to generate alerts in the ZTF (so they overlap the AGN lcc objects) and are thus comparable to the S82 QSO sample. Incidentally, the S82 QSO sources reach larger characteristic timescales than our objects. This is most likely due to a longer time span of the S82 light curves, which were taken in yearly seasons about 2–3 months long over the 2000-2010 decade. In fact, it has been demonstrated that the  DRW $\tau$ determination is biased for light curve lengths shorter than 10 times the true $\tau$ value \citep{2017koz,2017sanchez}, which is the case of our ZTF data.


Remarkably, all the variability features considered in this work distribute around different values for the Weak Type 1 (classified by a detailed spectroscopic analysis) and Type 2 SDSS samples. This implies the features can be used to statistically distinguish the obscured, Type 2 AGNs, from the variable, unobscured Type 1s even when the activity is weak. In particular, we tested the ability of the DRW parameters to separate the populations according to their variability. As a result, we confirmed that sources with DRW parameters that lie within the locus of Weak Type 1s in the DRW space show also a distribution of EW H$\alpha$ that skews towards larger values than the sources with lower variability amplitude and shorter timescales. We note that an overlap in the variability features is always present between these two samples, and could be occupied by intermediate type (1.8/1.9) AGNs that we have not considered in this work.  

 These encouraging results prove the ability of the optical variability to separate different populations. Following this idea, we can estimate the amount of variable objects in the Type 2 SDSS sample by applying limiting values in the variability features. Requiring all the variances to be positive (ExcessVar g$>0$ and MH $\sigma^2_{300,150g} >0 $) and the DRW parameters to be constrained within a certain range ($0.03 <$ SF$_\infty, g < 1$ mag, $ 1< \tau_g < 1000$ d) leads to 1361 variable sources ($\sim$11 per cent,  1361/12743). This value is comparable to the 10 per cent of variable sources found in a Type 2 sample in \citet{barth2014} (17/173) and 10 per cent of \textit{naked} AGNs reported by \citet{2004hawkins}.

By applying more restricting limits in the variability features, we searched for the most variable, Type 1-like objects in the whole sample to investigate whether they could be misclassified Type 1s or CL/CS AGNs. We found that just the $\sim$1 per cent of sources (246/20583) met the requirements, of which more than half were already classified as Type 1 in different catalogues. From the remaining 89 sources, we found 77 whose SDSS spectrum was already consistent with a Type 1, suggesting that variability selection is an efficient method for identifying Type 1 contaminants in Type 2 AGN samples, as also concluded by \citet{barth2014}. We note that 32 of the most variable sources present very weak BELs (5 <EW H$\alpha$ <15 \AA) and large variances, which could indicate a change in the accretion state of the sources, or in other words, they are CS candidates. Since several studies have reported extreme flux variability behaviour without significant spectroscopic changes \citep{2017graham,graham2020}, the CS nature of these sources needs to be verified through a new spectroscopic campaign.
 
Similarly, 12 of the most variable sources show no evidence of BELs in their SDSS spectra (EW H$\alpha$< 5\AA), which makes them potential CL/CS candidates. This selection method is very similar to the one applied in \citet{lopez2022}, where we confirmed new CL AGNs found with ALeRCE. In fact, 20 of their reported CL candidates are also within our most variable sources. Here, we improve the completeness of the candidate list by characterising light curves that use the available epochs of ZTF instead of only using the ZTF alert stream. This includes more of the lower-variability AGNs, which comprise 90 per cent of known Type 1 sources. This variability-based selection method (that is, searching for current Type-1 variability in spectrally classified Type 2 AGNs) has been claimed to be one of the most successful methods in the search of CL AGNs, with a success rate (SR) of $\sim 60$ per cent of CL confirmations \citep{lopez2022}. A spectroscopic confirmation is needed to determine the SR of the selection strategy presented in this work. In comparison, the highest SR to date is as high as 70 per cent, and has been reported by \citet{2021hon}. In the study, the authors searched for Type 1-like colours in a Type 2 sample coming from the spectroscopic Six-degree Field Galaxy Survey (6dFGS), which was observed $\sim$ 15 yr before.


\section{Summary and conclusions}
Although optical variability has been broadly used to characterize and separate different classes of objects, the variability in Type 2  AGNs has seldom been examined. In this work, we analyse systematically the ZTF light curves of $\sim$ 50 per cent (>15000) of AGNs from the SDSS DR16 to explore different variability features that allows to separate between obscured, Type 2 AGNs, from the variable, unobscured weak Type 1s. Our conclusions can be summarised as follows:
\begin{itemize}
    \item Our results indicate Type 2 AGNs show negligible optical variations, which is consistent with the general expectations from the simplest unified models. In comparison, the variances in the \textit{g} band are around 1 order of magnitude smaller than Type 1 objects with weak BELs.
    \item A small amount of variability features are able to separate distinct families of AGNs, including Type 1 and Type 2s AGNs. 
    \item The characteristic timescale $\tau$ and long term structure function SF$_{\infty,g}$ of the variations, given by the DRW model, are a powerful tool to separate weak Type 1 from Type 2 AGNs. Here, we find significantly higher EW of broad  H$\alpha$ for objects with $\tau _{g}> 15$ d and SF$_{\infty,g}>0.07$ mag.
    \item Around $\sim$11 per cent of Type 2 AGNs show evidence for optical variations, similarly to previous studies. This number of variable sources suggests a significant contamination of Type 1s or CL/CS AGNs in the DR16 AGN sample, that could be reduced by optical variability analysis.
    \item A spectroscopic analysis of the most variable Type 2 objects ($<$1 per cent, 89/12743) leads to the discovery of 77 weak Type 1 AGNs (EW H$\alpha$>5 \AA) and 12 CL/CS candidates (EW H$\alpha$<5 \AA). Follow-up spectroscopy would be needed to confirm the CL/CS nature of these sources and whether the weak Type 1s currently show larger BELs.
\end{itemize}
 
Future work with ZTF and next generation sky surveys such as the LSST, together with the use of machine learning algorithms, will allow to effectively improve the selection of pure samples of Type 2 AGNs in the optical range. This effort will be also crucial to better understand the CL/CS AGN population and the accretion physics at its most critical regimes, i.e., the advancing/receding accretion discs and the formation of the BLR clouds. 

\section*{Acknowledgements}

ELN and SB acknowledge support from Agencia Nacional de Investigación y Desarrollo / Programa de Becas/ Doctorado Nacional 21200718 and 21212344. ELN acknowledges the California Institute of Technology for its hospitality. PA, ELN and PL acknowledge financial support from Millenium Nucleus NCN$19\_058$ (TITANs). LHG acknowledges funds by ANID – Millennium Science Initiative Program – ICN12$\_$009 awarded to the Millennium Institute of Astrophysics (MAS). PL acknowledges partial support from FONDECYT through grant Nº 1201748. PSS acknowledges funds by ANID grant FONDECYT Postdoctorado Nº 3200250. MJG acknowledges partial support from the NSF grant AST-2108402. Based on observations collected at the Samuel Oschin Telescope 48-inch and the 60-inch Telescope at the Palomar Observatory as part of the Zwicky Transient Facility project. The ZTF forced-photometry service was funded under the Heising-Simons Foundation grant 12540303 (PI: Graham).

\section*{Data Availability}

The SDSS data underlying this article were accessed from SDSS DR16 (\url{http://skyserver.sdss.org/dr16}). The ZTF data underlying this article will be shared on reasonable request to the corresponding authors.



\bibliographystyle{mnras}
\bibliography{mnras_template} 

\begin{thebibliography}{}
\makeatletter
\relax
\def\mn@urlcharsother{\let\do\@makeother \do\$\do\&\do\#\do\^\do\_\do\%\do\~}
\def\mn@doi{\begingroup\mn@urlcharsother \@ifnextchar [ {\mn@doi@}
  {\mn@doi@[]}}
\def\mn@doi@[#1]#2{\def\@tempa{#1}\ifx\@tempa\@empty \href
  {http://dx.doi.org/#2} {doi:#2}\else \href {http://dx.doi.org/#2} {#1}\fi
  \endgroup}
\def\mn@eprint#1#2{\mn@eprint@#1:#2::\@nil}
\def\mn@eprint@arXiv#1{\href {http://arxiv.org/abs/#1} {{\tt arXiv:#1}}}
\def\mn@eprint@dblp#1{\href {http://dblp.uni-trier.de/rec/bibtex/#1.xml}
  {dblp:#1}}
\def\mn@eprint@#1:#2:#3:#4\@nil{\def\@tempa {#1}\def\@tempb {#2}\def\@tempc
  {#3}\ifx \@tempc \@empty \let \@tempc \@tempb \let \@tempb \@tempa \fi \ifx
  \@tempb \@empty \def\@tempb {arXiv}\fi \@ifundefined
  {mn@eprint@\@tempb}{\@tempb:\@tempc}{\expandafter \expandafter \csname
  mn@eprint@\@tempb\endcsname \expandafter{\@tempc}}}

\bibitem[\protect\citeauthoryear{{Ahumada} et~al.,}{{Ahumada}
  et~al.}{2020}]{DR16}
{Ahumada} R.,  et~al., 2020, \mn@doi [\apjs] {10.3847/1538-4365/ab929e}, \href
  {https://ui.adsabs.harvard.edu/abs/2020ApJS..249....3A} {249, 3}

\bibitem[\protect\citeauthoryear{{Antonucci}}{{Antonucci}}{1993}]{antonucci1993}
{Antonucci} R.,  1993, \mn@doi [\araa] {10.1146/annurev.aa.31.090193.002353},
  \href {https://ui.adsabs.harvard.edu/abs/1993ARA&A..31..473A} {31, 473}

\bibitem[\protect\citeauthoryear{{Ar{\'e}valo}, {Churazov}, {Zhuravleva},
  {Hern{\'a}ndez-Monteagudo}  \& {Revnivtsev}}{{Ar{\'e}valo}
  et~al.}{2012}]{Arevalo2012}
{Ar{\'e}valo} P.,  {Churazov} E.,  {Zhuravleva} I.,  {Hern{\'a}ndez-Monteagudo}
  C.,   {Revnivtsev} M.,  2012, \mn@doi [\mnras]
  {10.1111/j.1365-2966.2012.21789.x}, \href
  {https://ui.adsabs.harvard.edu/abs/2012MNRAS.426.1793A} {426, 1793}

\bibitem[\protect\citeauthoryear{{Baldwin}, {Phillips}  \&
  {Terlevich}}{{Baldwin} et~al.}{1981}]{1981bpt}
{Baldwin} J.~A.,  {Phillips} M.~M.,   {Terlevich} R.,  1981, \mn@doi [\pasp]
  {10.1086/130766}, \href
  {https://ui.adsabs.harvard.edu/abs/1981PASP...93....5B} {93, 5}

\bibitem[\protect\citeauthoryear{{Barth}, {Voevodkin}, {Carson}  \&
  {Wo{\'z}niak}}{{Barth} et~al.}{2014}]{barth2014}
{Barth} A.~J.,  {Voevodkin} A.,  {Carson} D.~J.,   {Wo{\'z}niak} P.,  2014,
  \mn@doi [\aj] {10.1088/0004-6256/147/1/12}, \href
  {https://ui.adsabs.harvard.edu/abs/2014AJ....147...12B} {147, 12}

\bibitem[\protect\citeauthoryear{{Bellm}}{{Bellm}}{2014}]{bellm2014}
{Bellm} E.,  2014, in {Wozniak} P.~R.,  {Graham} M.~J.,  {Mahabal} A.~A.,
  {Seaman} R.,  eds, The Third Hot-wiring the Transient Universe Workshop. pp
  27--33 (\mn@eprint {arXiv} {1410.8185})

\bibitem[\protect\citeauthoryear{{Bellm} et~al.,}{{Bellm}
  et~al.}{2019}]{bellm2019}
{Bellm} E.~C.,  et~al., 2019, \mn@doi [\pasp] {10.1088/1538-3873/aaecbe}, \href
  {https://ui.adsabs.harvard.edu/abs/2019PASP..131a8002B} {131, 018002}

\bibitem[\protect\citeauthoryear{{Bianchi}, {Corral}, {Panessa}, {Barcons},
  {Matt}, {Bassani}, {Carrera}  \& {Jim{\'e}nez-Bail{\'o}n}}{{Bianchi}
  et~al.}{2008}]{2008bianchi}
{Bianchi} S.,  {Corral} A.,  {Panessa} F.,  {Barcons} X.,  {Matt} G.,
  {Bassani} L.,  {Carrera} F.~J.,   {Jim{\'e}nez-Bail{\'o}n} E.,  2008, \mn@doi
  [\mnras] {10.1111/j.1365-2966.2007.12625.x}, \href
  {https://ui.adsabs.harvard.edu/abs/2008MNRAS.385..195B} {385, 195}

\bibitem[\protect\citeauthoryear{{Bolton} et~al.,}{{Bolton}
  et~al.}{2012}]{2012bolton}
{Bolton} A.~S.,  et~al., 2012, \mn@doi [\aj] {10.1088/0004-6256/144/5/144},
  \href {https://ui.adsabs.harvard.edu/abs/2012AJ....144..144B} {144, 144}

\bibitem[\protect\citeauthoryear{{Brightman} \& {Nandra}}{{Brightman} \&
  {Nandra}}{2008}]{2008brightman}
{Brightman} M.,  {Nandra} K.,  2008, \mn@doi [\mnras]
  {10.1111/j.1365-2966.2008.13841.x}, \href
  {https://ui.adsabs.harvard.edu/abs/2008MNRAS.390.1241B} {390, 1241}

\bibitem[\protect\citeauthoryear{{Cappellari}}{{Cappellari}}{2017}]{cappellari2017improving}
{Cappellari} M.,  2017, \mn@doi [\mnras] {10.1093/mnras/stw3020}, \href
  {https://ui.adsabs.harvard.edu/abs/2017MNRAS.466..798C} {466, 798}

\bibitem[\protect\citeauthoryear{{Chambers} \& {et al.}}{{Chambers} \& {et
  al.}}{2017}]{PS1}
{Chambers} K.~C.,  {et al.} 2017, VizieR Online Data Catalog, \href
  {https://ui.adsabs.harvard.edu/abs/2017yCat.2349....0C} {p. II/349}

\bibitem[\protect\citeauthoryear{{Dou} et~al.,}{{Dou} et~al.}{2022}]{2022dou}
{Dou} L.,  et~al., 2022, arXiv e-prints, \href
  {https://ui.adsabs.harvard.edu/abs/2022arXiv220811968D} {p. arXiv:2208.11968}

\bibitem[\protect\citeauthoryear{{Edelson}, {Turner}, {Pounds}, {Vaughan},
  {Markowitz}, {Marshall}, {Dobbie}  \& {Warwick}}{{Edelson}
  et~al.}{2002}]{2002edelson}
{Edelson} R.,  {Turner} T.~J.,  {Pounds} K.,  {Vaughan} S.,  {Markowitz} A.,
  {Marshall} H.,  {Dobbie} P.,   {Warwick} R.,  2002, \mn@doi [\apj]
  {10.1086/323779}, \href
  {https://ui.adsabs.harvard.edu/abs/2002ApJ...568..610E} {568, 610}

\bibitem[\protect\citeauthoryear{{Flesch}}{{Flesch}}{2021}]{2021flesch}
{Flesch} E.~W.,  2021, arXiv e-prints, \href
  {https://ui.adsabs.harvard.edu/abs/2021arXiv210512985F} {p. arXiv:2105.12985}

\bibitem[\protect\citeauthoryear{{F{\"o}rster} et~al.,}{{F{\"o}rster}
  et~al.}{2021}]{forster2021}
{F{\"o}rster} F.,  et~al., 2021, \mn@doi [\aj] {10.3847/1538-3881/abe9bc},
  \href {https://ui.adsabs.harvard.edu/abs/2021AJ....161..242F} {161, 242}

\bibitem[\protect\citeauthoryear{{Graham}, {Djorgovski}, {Drake}, {Stern},
  {Mahabal}, {Glikman}, {Larson}  \& {Christensen}}{{Graham}
  et~al.}{2017}]{2017graham}
{Graham} M.~J.,  {Djorgovski} S.~G.,  {Drake} A.~J.,  {Stern} D.,  {Mahabal}
  A.~A.,  {Glikman} E.,  {Larson} S.,   {Christensen} E.,  2017, \mn@doi
  [\mnras] {10.1093/mnras/stx1456}, \href
  {https://ui.adsabs.harvard.edu/abs/2017MNRAS.470.4112G} {470, 4112}

\bibitem[\protect\citeauthoryear{{Graham} et~al.,}{{Graham}
  et~al.}{2020}]{graham2020}
{Graham} M.~J.,  et~al., 2020, \mn@doi [\mnras] {10.1093/mnras/stz3244}, \href
  {https://ui.adsabs.harvard.edu/abs/2020MNRAS.491.4925G} {491, 4925}

\bibitem[\protect\citeauthoryear{{Guolo}, {Ruschel-Dutra}, {Grupe}, {Peterson},
  {Storchi-Bergmann}, {Schimoia}, {Nemmen}  \& {Robinson}}{{Guolo}
  et~al.}{2021}]{2021guolo}
{Guolo} M.,  {Ruschel-Dutra} D.,  {Grupe} D.,  {Peterson} B.~M.,
  {Storchi-Bergmann} T.,  {Schimoia} J.,  {Nemmen} R.,   {Robinson} A.,  2021,
  \mn@doi [\mnras] {10.1093/mnras/stab2550}, \href
  {https://ui.adsabs.harvard.edu/abs/2021MNRAS.508..144G} {508, 144}

\bibitem[\protect\citeauthoryear{{Hawkins}}{{Hawkins}}{2004}]{2004hawkins}
{Hawkins} M.~R.~S.,  2004, \mn@doi [\aap] {10.1051/0004-6361:20041127}, \href
  {https://ui.adsabs.harvard.edu/abs/2004A&A...424..519H} {424, 519}

\bibitem[\protect\citeauthoryear{Hon, Wolf, Onken, Webster  \& Auchettl}{Hon
  et~al.}{2021}]{2021hon}
Hon W.~J.,  Wolf C.,  Onken C.~A.,  Webster R.,   Auchettl K.,  2021, \mn@doi
  [\mnras] {10.1093/mnras/stab3694}, 511, 54

\bibitem[\protect\citeauthoryear{{Hutsem{\'e}kers}, {Ag{\'\i}s Gonz{\'a}lez},
  {Sluse}, {Ramos Almeida}  \& {Acosta Pulido}}{{Hutsem{\'e}kers}
  et~al.}{2017}]{hutsemekers2017}
{Hutsem{\'e}kers} D.,  {Ag{\'\i}s Gonz{\'a}lez} B.,  {Sluse} D.,  {Ramos
  Almeida} C.,   {Acosta Pulido} J.~A.,  2017, \mn@doi [\aap]
  {10.1051/0004-6361/201731397}, \href
  {https://ui.adsabs.harvard.edu/abs/2017A&A...604L...3H} {604, L3}

\bibitem[\protect\citeauthoryear{{Hutsem{\'e}kers}, {Ag{\'\i}s Gonz{\'a}lez},
  {Marin}, {Sluse}, {Ramos Almeida}  \& {Acosta Pulido}}{{Hutsem{\'e}kers}
  et~al.}{2019}]{hutsemekers2019}
{Hutsem{\'e}kers} D.,  {Ag{\'\i}s Gonz{\'a}lez} B.,  {Marin} F.,  {Sluse} D.,
  {Ramos Almeida} C.,   {Acosta Pulido} J.~A.,  2019, \mn@doi [\aap]
  {10.1051/0004-6361/201834633}, \href
  {https://ui.adsabs.harvard.edu/abs/2019A&A...625A..54H} {625, A54}

\bibitem[\protect\citeauthoryear{{Jiang} et~al.,}{{Jiang}
  et~al.}{2022}]{2022jiang}
{Jiang} N.,  et~al., 2022, arXiv e-prints, \href
  {https://ui.adsabs.harvard.edu/abs/2022arXiv220111633J} {p. arXiv:2201.11633}

\bibitem[\protect\citeauthoryear{{Kelly}, {Bechtold}  \&
  {Siemiginowska}}{{Kelly} et~al.}{2009}]{2009kelly}
{Kelly} B.~C.,  {Bechtold} J.,   {Siemiginowska} A.,  2009, \mn@doi [\apj]
  {10.1088/0004-637X/698/1/895}, \href
  {https://ui.adsabs.harvard.edu/abs/2009ApJ...698..895K} {698, 895}

\bibitem[\protect\citeauthoryear{{Koss} et~al.,}{{Koss}
  et~al.}{2017}]{2017koss}
{Koss} M.,  et~al., 2017, \mn@doi [\apj] {10.3847/1538-4357/aa8ec9}, \href
  {https://ui.adsabs.harvard.edu/abs/2017ApJ...850...74K} {850, 74}

\bibitem[\protect\citeauthoryear{{Koz{\l}owski}}{{Koz{\l}owski}}{2017}]{2017koz}
{Koz{\l}owski} S.,  2017, \mn@doi [\aap] {10.1051/0004-6361/201629890}, \href
  {https://ui.adsabs.harvard.edu/abs/2017A&A...597A.128K} {597, A128}

\bibitem[\protect\citeauthoryear{{LaMassa} et~al.,}{{LaMassa}
  et~al.}{2015}]{lamassa2015}
{LaMassa} S.~M.,  et~al., 2015, \mn@doi [\apj] {10.1088/0004-637X/800/2/144},
  \href {https://ui.adsabs.harvard.edu/abs/2015ApJ...800..144L} {800, 144}

\bibitem[\protect\citeauthoryear{{Liu}, {Liu}, {Dong}, {Zhou}, {Wang}, {Lu}  \&
  {Yuan}}{{Liu} et~al.}{2019}]{2019liu}
{Liu} H.-Y.,  {Liu} W.-J.,  {Dong} X.-B.,  {Zhou} H.,  {Wang} T.,  {Lu} H.,
  {Yuan} W.,  2019, \mn@doi [\apjs] {10.3847/1538-4365/ab298b}, \href
  {https://ui.adsabs.harvard.edu/abs/2019ApJS..243...21L} {243, 21}

\bibitem[\protect\citeauthoryear{{L{\'o}pez-Navas} et~al.,}{{L{\'o}pez-Navas}
  et~al.}{2022}]{lopez2022}
{L{\'o}pez-Navas} E.,  et~al., 2022, \mn@doi [\mnras] {10.1093/mnrasl/slac033},
  \href {https://ui.adsabs.harvard.edu/abs/2022MNRAS.513L..57L} {513, L57}

\bibitem[\protect\citeauthoryear{{MacLeod} et~al.,}{{MacLeod}
  et~al.}{2010}]{2010macleod}
{MacLeod} C.~L.,  et~al., 2010, \mn@doi [\apj] {10.1088/0004-637X/721/2/1014},
  \href {https://ui.adsabs.harvard.edu/abs/2010ApJ...721.1014M} {721, 1014}

\bibitem[\protect\citeauthoryear{{MacLeod} et~al.,}{{MacLeod}
  et~al.}{2016}]{macleod2016}
{MacLeod} C.~L.,  et~al., 2016, \mn@doi [\mnras] {10.1093/mnras/stv2997}, \href
  {https://ui.adsabs.harvard.edu/abs/2016MNRAS.457..389M} {457, 389}

\bibitem[\protect\citeauthoryear{{Masci} et~al.,}{{Masci}
  et~al.}{2019}]{masci2019}
{Masci} F.~J.,  et~al., 2019, \mn@doi [\pasp] {10.1088/1538-3873/aae8ac}, \href
  {https://ui.adsabs.harvard.edu/abs/2019PASP..131a8003M} {131, 018003}

\bibitem[\protect\citeauthoryear{{Massaro}, {Maselli}, {Leto}, {Marchegiani},
  {Perri}, {Giommi}  \& {Piranomonte}}{{Massaro} et~al.}{2015}]{2015massaro}
{Massaro} E.,  {Maselli} A.,  {Leto} C.,  {Marchegiani} P.,  {Perri} M.,
  {Giommi} P.,   {Piranomonte} S.,  2015, \mn@doi [\apss]
  {10.1007/s10509-015-2254-2}, \href
  {https://ui.adsabs.harvard.edu/abs/2015Ap&SS.357...75M} {357, 75}

\bibitem[\protect\citeauthoryear{{Nandra}, {George}, {Mushotzky}, {Turner}  \&
  {Yaqoob}}{{Nandra} et~al.}{1997}]{1997nandra}
{Nandra} K.,  {George} I.~M.,  {Mushotzky} R.~F.,  {Turner} T.~J.,   {Yaqoob}
  T.,  1997, \mn@doi [\apj] {10.1086/303600}, \href
  {https://ui.adsabs.harvard.edu/abs/1997ApJ...476...70N} {476, 70}

\bibitem[\protect\citeauthoryear{{Noda} \& {Done}}{{Noda} \&
  {Done}}{2018}]{Noda2018}
{Noda} H.,  {Done} C.,  2018, \mn@doi [\mnras] {10.1093/mnras/sty2032}, \href
  {https://ui.adsabs.harvard.edu/abs/2018MNRAS.480.3898N} {480, 3898}

\bibitem[\protect\citeauthoryear{{Oh}, {Yi}, {Schawinski}, {Koss},
  {Trakhtenbrot}  \& {Soto}}{{Oh} et~al.}{2015}]{Oh2015}
{Oh} K.,  {Yi} S.~K.,  {Schawinski} K.,  {Koss} M.,  {Trakhtenbrot} B.,
  {Soto} K.,  2015, \mn@doi [\apjs] {10.1088/0067-0049/219/1/1}, \href
  {https://ui.adsabs.harvard.edu/abs/2015ApJS..219....1O} {219, 1}

\bibitem[\protect\citeauthoryear{{Panessa} et~al.,}{{Panessa}
  et~al.}{2009}]{2009panessa}
{Panessa} F.,  et~al., 2009, \mn@doi [\mnras]
  {10.1111/j.1365-2966.2009.15225.x}, \href
  {https://ui.adsabs.harvard.edu/abs/2009MNRAS.398.1951P} {398, 1951}

\bibitem[\protect\citeauthoryear{{Pappa}, {Georgantopoulos}, {Stewart}  \&
  {Zezas}}{{Pappa} et~al.}{2001}]{2001Pappa}
{Pappa} A.,  {Georgantopoulos} I.,  {Stewart} G.~C.,   {Zezas} A.~L.,  2001,
  \mn@doi [\mnras] {10.1046/j.1365-8711.2001.04609.x}, \href
  {https://ui.adsabs.harvard.edu/abs/2001MNRAS.326..995P} {326, 995}

\bibitem[\protect\citeauthoryear{{Predehl} \& {Schmitt}}{{Predehl} \&
  {Schmitt}}{1995}]{1995predehl}
{Predehl} P.,  {Schmitt} J.~H.~M.~M.,  1995, \aap, \href
  {https://ui.adsabs.harvard.edu/abs/1995A&A...293..889P} {293, 889}

\bibitem[\protect\citeauthoryear{{Rivers} et~al.,}{{Rivers}
  et~al.}{2015}]{2015rivers}
{Rivers} E.,  et~al., 2015, \mn@doi [\apj] {10.1088/0004-637X/815/1/55}, \href
  {https://ui.adsabs.harvard.edu/abs/2015ApJ...815...55R} {815, 55}

\bibitem[\protect\citeauthoryear{{Runnoe} et~al.,}{{Runnoe}
  et~al.}{2016}]{runnoe2016}
{Runnoe} J.~C.,  et~al., 2016, \mn@doi [\mnras] {10.1093/mnras/stv2385}, \href
  {https://ui.adsabs.harvard.edu/abs/2016MNRAS.455.1691R} {455, 1691}

\bibitem[\protect\citeauthoryear{{S{\'a}nchez-S{\'a}ez}
  et~al.,}{{S{\'a}nchez-S{\'a}ez} et~al.}{2021}]{sanchez2021}
{S{\'a}nchez-S{\'a}ez} P.,  et~al., 2021, \mn@doi [\aj]
  {10.3847/1538-3881/abd5c1}, \href
  {https://ui.adsabs.harvard.edu/abs/2021AJ....161..141S} {161, 141}

\bibitem[\protect\citeauthoryear{{S{\'a}nchez} et~al.,}{{S{\'a}nchez}
  et~al.}{2017}]{2017sanchez}
{S{\'a}nchez} P.,  et~al., 2017, \mn@doi [\apj] {10.3847/1538-4357/aa9188},
  \href {https://ui.adsabs.harvard.edu/abs/2017ApJ...849..110S} {849, 110}

\bibitem[\protect\citeauthoryear{{Sheng}, {Wang}, {Jiang}, {Yang}, {Yan}, {Dou}
   \& {Peng}}{{Sheng} et~al.}{2017}]{2017sheng}
{Sheng} Z.,  {Wang} T.,  {Jiang} N.,  {Yang} C.,  {Yan} L.,  {Dou} L.,   {Peng}
  B.,  2017, \mn@doi [\apjl] {10.3847/2041-8213/aa85de}, \href
  {https://ui.adsabs.harvard.edu/abs/2017ApJ...846L...7S} {846, L7}

\bibitem[\protect\citeauthoryear{{Shi}, {Rieke}, {Smith}, {Rigby}, {Hines},
  {Donley}, {Schmidt}  \& {Diamond-Stanic}}{{Shi} et~al.}{2010}]{2010Shi}
{Shi} Y.,  {Rieke} G.~H.,  {Smith} P.,  {Rigby} J.,  {Hines} D.,  {Donley} J.,
  {Schmidt} G.,   {Diamond-Stanic} A.~M.,  2010, \mn@doi [\apj]
  {10.1088/0004-637X/714/1/115}, \href
  {https://ui.adsabs.harvard.edu/abs/2010ApJ...714..115S} {714, 115}

\bibitem[\protect\citeauthoryear{{Trakhtenbrot} et~al.,}{{Trakhtenbrot}
  et~al.}{2019}]{trakhtenbrot2019}
{Trakhtenbrot} B.,  et~al., 2019, \mn@doi [\apj] {10.3847/1538-4357/ab39e4},
  \href {https://ui.adsabs.harvard.edu/abs/2019ApJ...883...94T} {883, 94}

\bibitem[\protect\citeauthoryear{{Tran}, {Lyke}  \& {Mader}}{{Tran}
  et~al.}{2011}]{2011tran}
{Tran} H.~D.,  {Lyke} J.~E.,   {Mader} J.~A.,  2011, \mn@doi [\apjl]
  {10.1088/2041-8205/726/2/L21}, \href
  {https://ui.adsabs.harvard.edu/abs/2011ApJ...726L..21T} {726, L21}

\bibitem[\protect\citeauthoryear{{Vaughan}, {Edelson}, {Warwick}  \&
  {Uttley}}{{Vaughan} et~al.}{2003}]{2003vaughan}
{Vaughan} S.,  {Edelson} R.,  {Warwick} R.~S.,   {Uttley} P.,  2003, \mn@doi
  [\mnras] {10.1046/j.1365-2966.2003.07042.x}, \href
  {https://ui.adsabs.harvard.edu/abs/2003MNRAS.345.1271V} {345, 1271}

\bibitem[\protect\citeauthoryear{{Vazdekis}, {Koleva}, {Ricciardelli},
  {R{\"o}ck}  \& {Falc{\'o}n-Barroso}}{{Vazdekis}
  et~al.}{2016}]{2016MNRAS.463.3409V}
{Vazdekis} A.,  {Koleva} M.,  {Ricciardelli} E.,  {R{\"o}ck} B.,
  {Falc{\'o}n-Barroso} J.,  2016, \mn@doi [\mnras] {10.1093/mnras/stw2231},
  \href {https://ui.adsabs.harvard.edu/abs/2016MNRAS.463.3409V} {463, 3409}

\bibitem[\protect\citeauthoryear{{Wang}, {Xu}, {Wang}, {Zhang}, {Zheng}  \&
  {Wei}}{{Wang} et~al.}{2019}]{2019wang}
{Wang} J.,  {Xu} D.~W.,  {Wang} Y.,  {Zhang} J.~B.,  {Zheng} J.,   {Wei} J.~Y.,
   2019, \mn@doi [\apj] {10.3847/1538-4357/ab4d90}, \href
  {https://ui.adsabs.harvard.edu/abs/2019ApJ...887...15W} {887, 15}

\bibitem[\protect\citeauthoryear{{Wolter}, {Gioia}, {Henry}  \&
  {Mullis}}{{Wolter} et~al.}{2005}]{2005wolter}
{Wolter} A.,  {Gioia} I.~M.,  {Henry} J.~P.,   {Mullis} C.~R.,  2005, \mn@doi
  [\aap] {10.1051/0004-6361:20053441}, \href
  {https://ui.adsabs.harvard.edu/abs/2005A&A...444..165W} {444, 165}

\bibitem[\protect\citeauthoryear{{Xia}, {Xue}, {Mao}, {Boller}, {Deng}  \&
  {Wu}}{{Xia} et~al.}{2002}]{2002xia}
{Xia} X.~Y.,  {Xue} S.~J.,  {Mao} S.,  {Boller} T.,  {Deng} Z.~G.,   {Wu} H.,
  2002, \mn@doi [\apj] {10.1086/324187}, \href
  {https://ui.adsabs.harvard.edu/abs/2002ApJ...564..196X} {564, 196}

\bibitem[\protect\citeauthoryear{{Yang}, {Shen}, {Liu}, {Wu}, {Jiang},
  {Shangguan}, {Graham}  \& {Yao}}{{Yang} et~al.}{2019}]{2019Yang}
{Yang} Q.,  {Shen} Y.,  {Liu} X.,  {Wu} X.-B.,  {Jiang} L.,  {Shangguan} J.,
  {Graham} M.~J.,   {Yao} S.,  2019, \mn@doi [\apj] {10.3847/1538-4357/ab481a},
  \href {https://ui.adsabs.harvard.edu/abs/2019ApJ...885..110Y} {885, 110}

\bibitem[\protect\citeauthoryear{{Yip} et~al.,}{{Yip} et~al.}{2009}]{2009yip}
{Yip} C.~W.,  et~al., 2009, \mn@doi [\aj] {10.1088/0004-6256/137/6/5120}, \href
  {https://ui.adsabs.harvard.edu/abs/2009AJ....137.5120Y} {137, 5120}

\bibitem[\protect\citeauthoryear{{Yuan}, {Strauss}  \& {Zakamska}}{{Yuan}
  et~al.}{2016}]{2016yuan}
{Yuan} S.,  {Strauss} M.~A.,   {Zakamska} N.~L.,  2016, \mn@doi [\mnras]
  {10.1093/mnras/stw1747}, \href
  {https://ui.adsabs.harvard.edu/abs/2016MNRAS.462.1603Y} {462, 1603}

\makeatother
\end{thebibliography}


\appendix

\section{Most variable sources} \label{appendix}
\subsection{Anomalous AGNs}
\begin{figure}
\centering
  \includegraphics[width=0.5\columnwidth]{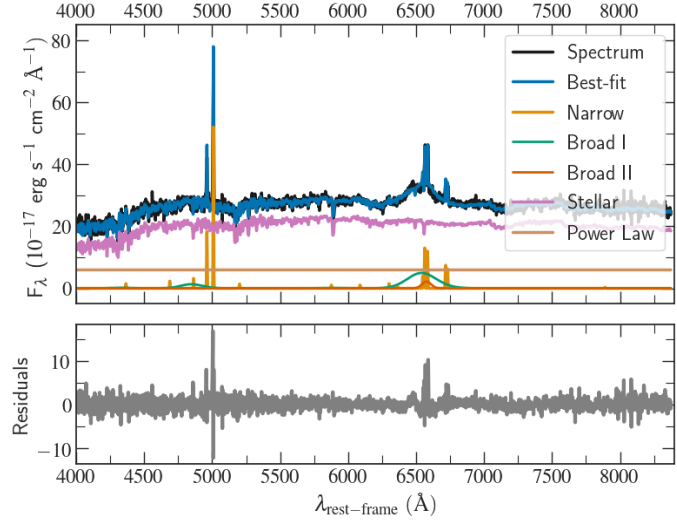}
 \caption{Fit to the SDSS spectrum of one of the most variable AGNs in our Type 2 SDSS sample, J1246+2820. Some residuals remain around the H$\alpha$ line as a result of a more complex profile-- notably a second peak of H$\alpha$ redshifted with respect to the narrow lines.  }

 \label{fig:anomalousspec}
\end{figure}

\begin{figure}
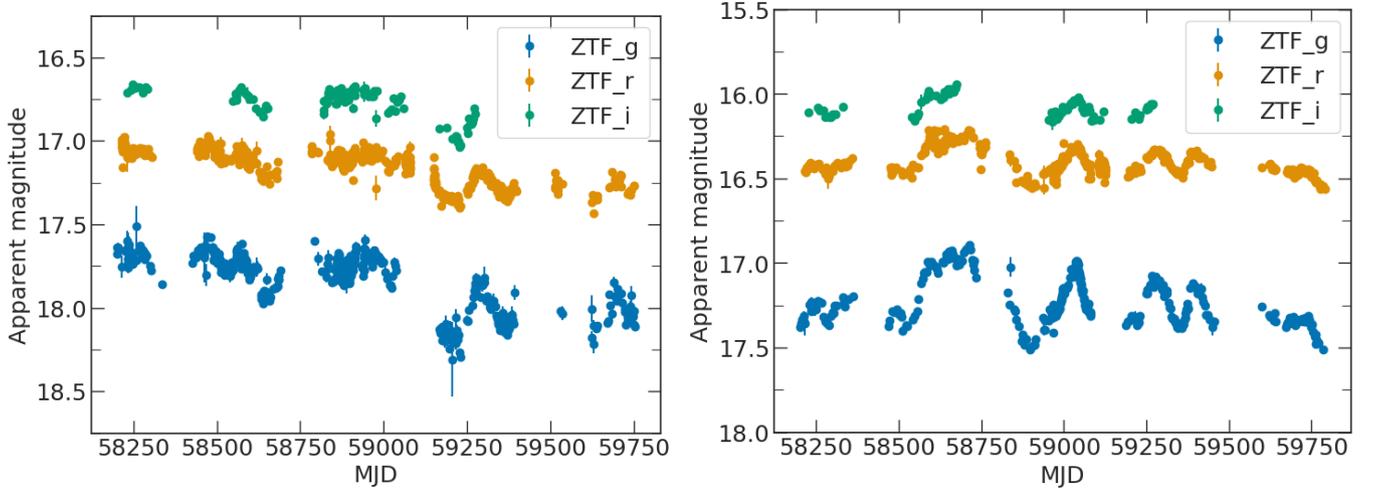


  \includegraphics[width=0.5\columnwidth]{plots/forced_phot1.pdf}
  \includegraphics[width=0.5\columnwidth]{plots/forced_phot2.pdf}

 \caption{ZTF light curves of J1246+2820 (left) and the SMBHB candidate host J1430+2303 (right). }
 
 \label{fig:anomalous}
\end{figure}

Within the most variable sources, we found three cases where two broad component where needed to fit the permitted emission lines in the SDSS spectrum: J124617.34+282033.92 (J1246+2820), J073149.29+361353.03 (J0731+3613) and J154246.71+334602.62 (J1542+3346). In these cases, the EW H$\alpha$ is computed by adding the flux of the two components. For J0731+3613 and J1542+3346, the addition of a second semi-broad component with a similar line-of-sight velocity to the broad one was sufficient to fit the spectral shape of the  H$\alpha$ emission.
On the contrary, for J1246+2820 the addition of a second broad component leads to some residuals around the H$\alpha$ region (within the standard deviation), as a result of a more complex profile. In Figure~ \ref{fig:anomalousspec} we present the fit to the SDSS spectrum of J1246+2820, where there seem to be a second peak of H$\alpha$ redshifted with respect to the narrow emission lines. Interestingly, J1246+2820 (full) ZTF light curve is similar to that from the nearby Seyfert galaxy SDSS J143016.05+230344.4 (J1430+2303, see Figure~ \ref{fig:anomalous} for the most updated light curves retrieved from the ZTF Forced Photometry service), which has been recently reported as a candidate to host a supermassive black hole binary \citep[SMBHB,][]{2022jiang,2022dou}. Its most recent optical spectrum, obtained on January 2022, also shows a complex velocity structure, which can be fitted with three Gaussians, including a significantly redshifted and a blueshifted component \citep{2022jiang}. This suggests that similar mechanisms could be producing the observed optical properties, but future multi-wavelength observations are needed to better understand the nature of this source.
\onecolumn

\begin{longtable}{ccccccccc}
    \caption{SDSS spectral information and fits for the most variable sources selected in this work. For the equivalent widths (EW) of H$\alpha$ and H$\beta$ we present the $_{10}^{90}$-th percentiles. Sources with EW H$\alpha$>5 \AA~ are considered weak Type 1 AGNs, while sources EW H$\alpha$<5 \AA~ are CL candidates. Names in bold refer to sources fitted with two broad components.  }
	\label{tab: fits}
\\
Name & Ra & Dec & redshift & MJD & fiberid & plate & EW H$\alpha$ & EW H$\beta$ \\
& deg & deg & & & & & \AA &\AA\\
    \hline
J000722.15+153811.56 & 1.842292 & 15.636544 & 0.12 & 52251 & 503 & 751 & 14$^{15}_{13} $ & 0$^{1}_{0}$\\
J002744.34+131300.22 & 6.93475 & 13.216728 & 0.09 & 56210 & 880 & 6190 & 21$^{23}_{20} $ & 0$^{1}_{0}$\\
J004247.82+231442.60 & 10.69925 & 23.245167 & 0.14 & 56248 & 300 & 6285 & 8$^{13}_{5} $ & 0$^{0}_{0}$\\
J005505.43+175256.18 & 13.772625 & 17.882272 & 0.1 & 56901 & 367 & 7623 & 21$^{23}_{20} $ & 1$^{1}_{0}$\\
J011142.76+262156.17 & 17.928167 & 26.365603 & 0.08 & 57367 & 852 & 7679 & 15$^{16}_{14} $ & 0$^{1}_{0}$\\
J011319.27+233449.13 & 18.330292 & 23.580314 & 0.11 & 55953 & 162 & 5699 & 5$^{5}_{4} $ & 0$^{0}_{0}$\\
J020858.95+274633.02 & 32.245625 & 27.775839 & 0.23 & 56325 & 320 & 6272 & 1$^{22}_{7} $ & 0$^{0}_{0}$\\
\textbf{J073149.29+361353.03} & 112.955375 & 36.231397 & 0.14 & 55182 & 447 & 3662 & 53$^{55}_{41} $ & 0$^{1}_{0}$\\
J081917.51+301935.76 & 124.822958 & 30.3266 & 0.10 & 52619 & 94 & 931 & 3$^{13}_{3} $ & 0$^{1}_{0}$\\
J082250.42+154025.89 & 125.710083 & 15.673858 & 0.12 & 53713 & 517 & 2272 & 26$^{28}_{23} $ & 0$^{1}_{0}$\\
J082517.65+125855.07 & 126.323542 & 12.981964 & 0.17 & 54096 & 593 & 2422 & 29$^{31}_{26} $ & 0$^{2}_{0}$\\
J083224.29+355135.92 & 128.101208 & 35.859978 & 0.14 & 52668 & 459 & 1197 & 14$^{18}_{12} $ & 0$^{1}_{0}$\\
J083310.47+041036.73 & 128.293625 & 4.176869 & 0.10 & 52646 & 190 & 1186 & 20$^{20}_{13} $ & 0$^{1}_{0}$\\
J083934.03+104925.07 & 129.891792 & 10.823631 & 0.18 & 54061 & 437 & 2573 & 3$^{5}_{2} $ & 0$^{1}_{0}$\\
J091214.33+065722.28 & 138.059708 & 6.956189 & 0.13 & 52703 & 479 & 1194 & 11$^{22}_{9} $ & 0$^{2}_{0}$\\
J091357.17+250813.95 & 138.488208 & 25.137208 & 0.18 & 53415 & 217 & 2087 & 21$^{23}_{19} $ & 0$^{2}_{0}$\\
J093612.22+253226.10 & 144.050917 & 25.540583 & 0.13 & 54524 & 592 & 2294 & 0$^{2}_{0} $ & 0$^{0}_{0}$\\
J095137.27+341612.30 & 147.905292 & 34.270083 & 0.13 & 53388 & 351 & 1948 & 12$^{26}_{10} $ & 0$^{2}_{0}$\\
J095504.39+072606.93 & 148.768292 & 7.435258 & 0.21 & 52734 & 140 & 1235 & 15$^{17}_{13} $ & 0$^{1}_{0}$\\
J095526.28+305057.86 & 148.8595 & 30.849406 & 0.09 & 53436 & 342 & 1950 & 8$^{11}_{8} $ & 0$^{1}_{0}$\\
J102935.82+244639.38 & 157.39925 & 24.777606 & 0.11 & 53734 & 545 & 2349 & 14$^{166}_{0} $ & 0$^{2}_{0}$\\
J103450.93+462915.31 & 158.712208 & 46.487586 & 0.18 & 52620 & 294 & 962 & $8^{9}_{3} $ & $0^{1.6}_{0}$ \\
J103812.23+244004.77 & 159.550958 & 24.667992 & 0.09 & 53770 & 150 & 2352 & 6$^{7}_{5} $ & 0$^{1}_{0}$\\
J110407.10+125005.21 & 166.029583 & 12.834781 & 0.13 & 53119 & 534 & 1603 & 3$^{5}_{2} $ & 0$^{1}_{0}$\\
J110529.70+051649.23 & 166.37375 & 5.280342 & 0.09 & 52356 & 400 & 581 & 12$^{18}_{11} $ & 0$^{0}_{0}$\\
J110920.32+531426.20 & 167.334667 & 53.240611 & 0.18 & 57135 & 363 & 8171 & 28$^{31}_{27} $ & 6$^{7}_{5}$\\
J111625.35+220049.37 & 169.105625 & 22.013714 & 0.14 & 54178 & 585 & 2492 & 39$^{52}_{39} $ & 0$^{8}_{0}$\\
J112229.25+102126.76 & 170.621875 & 10.357433 & 0.21 & 55976 & 584 & 5371 & 13$^{21}_{13} $ & 0$^{2}_{0}$\\
J112634.19+395539.72 & 171.642458 & 39.9277 & 0.19 & 53436 & 495 & 1996 & 38$^{51}_{35} $ & 1$^{5}_{0}$\\
J112939.08+365300.97 & 172.412833 & 36.883603 & 0.20 & 55673 & 426 & 4648 & 44$^{84}_{40} $ & 2$^{10}_{0}$\\
J114931.03+163743.16 & 177.379292 & 16.628656 & 0.29 & 56035 & 35 & 5892 & 31$^{34}_{27} $ & 4$^{6}_{3}$\\
J120045.49+145803.67 & 180.189542 & 14.967686 & 0.11 & 53463 & 158 & 1763 & 10$^{21}_{7} $ & 0$^{1}_{0}$\\
J120054.50+145850.97 & 180.227083 & 14.980825 & 0.08 & 53463 & 595 & 1763 & 4$^{11}_{4} $ & 0$^{0}_{0}$\\
J120349.21+605317.45 & 180.955042 & 60.888181 & 0.07 & 52405 & 420 & 954 & 24$^{26}_{23} $ & 0$^{1}_{0}$\\
J120402.12+335247.47 & 181.008833 & 33.879853 & 0.14 & 53469 & 38 & 2099 & 26$^{49}_{24} $ & 0$^{6}_{0}$\\
J120459.00+153513.85 & 181.245833 & 15.587181 & 0.22 & 53467 & 447 & 1764 & 25$^{27}_{21} $ & 0$^{2}_{0}$\\
J122011.98+153029.97 & 185.049917 & 15.508325 & 0.22 & 53436 & 382 & 1767 & 28$^{34}_{24} $ & 2$^{5}_{0}$\\
J122026.76+363327.88 & 185.1115 & 36.557744 & 0.10 & 53442 & 423 & 2003 & 20$^{21}_{19} $ & 0$^{1}_{0}$\\
J122053.47+283239.80 & 185.222792 & 28.544389 & 0.09 & 53816 & 569 & 2231 & 24$^{25}_{22} $ & 3$^{4}_{2}$\\
J122415.44+272506.48 & 186.064333 & 27.418467 & 0.09 & 56356 & 786 & 5976 & 22$^{23}_{21} $ & 4$^{5}_{4}$\\
J122439.67+180630.93 & 186.165292 & 18.108592 & 0.13 & 56034 & 566 & 5852 & 21$^{22}_{19} $ & 1$^{2}_{0}$\\
J122737.79+084406.84 & 186.907458 & 8.735233 & 0.09 & 53472 & 566 & 1626 & 12$^{19}_{12} $ & 0$^{0}_{0}$\\
J122801.79+223200.99 & 187.007458 & 22.533608 & 0.16 & 54495 & 355 & 2647 & $29^{31}_{27} $ & $0^{1}_{0}$\\
J124214.48+141146.99 & 190.560333 & 14.196386 & 0.16 & 53502 & 59 & 1769 & 15$^{17}_{13} $ & 0$^{1}_{0}$\\
J124445.48+282958.63 & 191.1895 & 28.499619 & 0.20 & 54205 & 369 & 2238 & 50$^{57}_{47} $ & 13$^{14}_{12}$\\
\textbf{J124617.34+282033.92} & 191.57225 & 28.342756 & 0.10 & 54205 & 442 & 2238 & 57$^{59}_{54} $ & 10$^{11}_{8}$\\
J124629.36+465932.76 & 191.622333 & 46.992433 & 0.23 & 53089 & 118 & 1455 & 27$^{29}_{26} $ & 0$^{2}_{0}$\\
J131130.66+315200.81 & 197.87775 & 31.866892 & 0.07 & 53819 & 632 & 2029 & 8$^{9}_{8} $ & 0$^{0}_{0}$\\
J131508.78+365334.61 & 198.786583 & 36.892947 & 0.27 & 57519 & 836 & 8871 & 5$^{6}_{3} $ & 0$^{1}_{0}$\\
J131858.85+573006.65 & 199.745208 & 57.501847 & 0.10 & 52759 & 338 & 1320 & 13$^{22}_{12} $ & 0$^{1}_{0}$\\
J132558.71+151257.99 & 201.494625 & 15.216108 & 0.20 & 53759 & 449 & 1774 & 17$^{29}_{16} $ & 0$^{1}_{0}$\\
J132939.91+323144.28 & 202.416292 & 32.528967 & 0.24 & 58526 & 146 & 10257 & $36^{40}_{30} $ & $9^{10}_{5}$\\
J133052.10+202600.99 & 202.717083 & 20.433608 & 0.22 & 54230 & 212 & 2653 & 19$^{22}_{18} $ & 6$^{8}_{4}$\\
J134045.31+405333.45 & 205.188792 & 40.892625 & 0.09 & 57511 & 440 & 8391 & 24$^{25}_{22} $ & 3$^{3}_{2}$\\
J134148.78+370047.13 & 205.45325 & 37.013092 & 0.20 & 53858 & 513 & 2101 & 12$^{33}_{10} $ & 0$^{4}_{0}$\\
J134803.73+453728.47 & 207.015542 & 45.624575 & 0.16 & 53082 & 31 & 1465 & $13^{24}_{13} $ & $0^{3}_{0}$\\
J135007.70+124657.41 & 207.532083 & 12.782614 & 0.14 & 53857 & 205 & 1777 & 22$^{28}_{21} $ & 2$^{5}_{2}$\\
J135425.48+334254.98 & 208.606167 & 33.715272 & 0.25 & 55274 & 746 & 3861 & 8$^{38}_{5} $ & 0$^{3}_{0}$\\
J141821.98+612731.79 & 214.591583 & 61.458831 & 0.25 & 52365 & 129 & 606 & 11$^{13}_{10} $ & 0$^{1}_{0}$\\
J142052.22+472625.72 & 215.217583 & 47.440478 & 0.13 & 53462 & 303 & 1673 & 4$^{5}_{3} $ & 0$^{1}_{0}$\\
J142352.09+245417.14 & 215.967042 & 24.904761 & 0.07 & 53493 & 417 & 2132 & 20$^{33}_{19} $ & 0$^{2}_{0}$\\
J142736.37+265700.50 & 216.901542 & 26.950139 & 0.17 & 53876 & 208 & 2134 & 12$^{20}_{9} $ & 1$^{3}_{0}$\\
J143519.06+511739.78 & 218.829417 & 51.294383 & 0.08 & 52781 & 245 & 1327 & 5$^{7}_{4} $ & 0$^{1}_{0}$\\
J144021.49+141125.74 & 220.089542 & 14.190483 & 0.12 & 54234 & 147 & 2748 & 13$^{15}_{11} $ & 0$^{1}_{0}$\\
J144933.48+082355.69 & 222.3895 & 8.398803 & 0.12 & 54555 & 418 & 1814 & 8$^{10}_{7} $ & 0$^{1}_{0}$\\
J145005.11+154348.27 & 222.521292 & 15.730075 & 0.27 & 54535 & 308 & 2764 & 24$^{30}_{21} $ & 2$^{4}_{0}$\\
J150111.92+040422.87 & 225.299667 & 4.073019 & 0.16 & 52055 & 478 & 589 & 8$^{29}_{5} $ & 0$^{1}_{0}$\\
J153006.52+071020.17 & 232.527167 & 7.172269 & 0.13 & 54208 & 322 & 1820 & 25$^{42}_{23} $ & 0$^{3}_{0}$\\
J153832.66+460735.01 & 234.636083 & 46.126392 & 0.2 & 52781 & 559 & 1332 & 13$^{16}_{11} $ & 0$^{1}_{0}$\\
\textbf{J154246.71+334602.62} & 235.694625 & 33.767394 & 0.28 & 55747 & 816 & 4971 & 95$^{102}_{93} $ & 17$^{19}_{15}$\\
J154907.53+372900.59 & 237.281375 & 37.483497 & 0.20 & 53172 & 270 & 1681 & 14$^{26}_{16} $ & 0$^{1}_{0}$\\
J155258.30+273728.41 & 238.242917 & 27.624558 & 0.09 & 53498 & 603 & 1654 & 25$^{27}_{24} $ & 0$^{1}_{0}$\\
J155259.94+210246.89 & 238.24975 & 21.046358 & 0.17 & 53557 & 220 & 2171 & 28$^{63}_{25} $ & 0$^{6}_{0}$\\
J155640.32+451338.41 & 239.168 & 45.227336 & 0.18 & 52753 & 471 & 1169 & 14$^{52}_{10} $ & 0$^{5}_{0}$\\
J161123.42+424139.79 & 242.847583 & 42.694386 & 0.25 & 57896 & 953 & 8528 & 0$^{29}_{0} $ & 0$^{4}_{0}$\\
J161219.56+462942.62 & 243.0815 & 46.495172 & 0.13 & 52443 & 414 & 814 & 4$^{22}_{4} $ & 0$^{2}_{0}$\\
J163344.96+112611.59 & 248.437333 & 11.436553 & 0.19 & 54585 & 212 & 2533 & 20$^{21}_{18} $ & 3$^{4}_{2}$\\
J163639.58+194201.73 & 249.164917 & 19.700481 & 0.15 & 53224 & 51 & 1659 & 8$^{29}_{7} $ & 0$^{1}_{0}$\\
J164360.00+321009.80 & 251.0 & 32.169389 & 0.14 & 52786 & 159 & 1341 & 22$^{24}_{20} $ & 0$^{1}_{0}$\\
J164432.94+213306.45 & 251.13725 & 21.551792 & 0.3 & 53149 & 80 & 1570 & 25$^{30}_{23} $ & 1$^{2}_{0}$\\
J165119.25+242011.41 & 252.830208 & 24.336503 & 0.15 & 52912 & 455 & 1424 & 2$^{23}_{3} $ & 0$^{3}_{0}$\\
J220138.03+121456.52 & 330.408458 & 12.249033 & 0.19 & 52224 & 520 & 734 & 5$^{36}_{0} $ & 0$^{3}_{0}$\\
J221044.76+245958.05 & 332.6865 & 24.999458 & 0.12 & 56213 & 258 & 5958 & 42$^{46}_{33} $ & 0$^{1}_{0}$\\
J222559.67+201944.75 & 336.498625 & 20.329097 & 0.16 & 55854 & 712 & 5024 & 20$^{21}_{19} $ & 1$^{2}_{0}$\\
J223249.33+035829.19 & 338.205542 & 3.974775 & 0.14 & 55525 & 402 & 4291 & 27$^{29}_{26} $ & 1$^{1}_{0}$\\
J231009.82+074928.51 & 347.540917 & 7.824586 & 0.16 & 56187 & 642 & 6168 & 52$^{53}_{51} $ & 0$^{2}_{0}$\\
J231138.89+274504.30 & 347.912042 & 27.751194 & 0.12 & 56559 & 358 & 6289 & 6$^{7}_{5} $ & 0$^{0}_{0}$\\
J231720.14+143855.97 & 349.333917 & 14.648881 & 0.15 & 52258 & 315 & 745 & 9$^{13}_{9} $ & 0$^{1}_{0}$\\
J233628.86+231956.29 & 354.12025 & 23.332303 & 0.18 & 56566 & 856 & 6519 & 8$^{15}_{7} $ & 0$^{0}_{0}$\\
\end{longtable}

\bsp	
\label{lastpage}
\end{document}